\documentclass[longauth]{aa}
\usepackage[varg]{txfonts}
\usepackage{graphicx}
\usepackage{natbib}
\usepackage{mathrsfs}
\usepackage{xcolor}
\bibpunct{(}{)}{;}{a}{}{,}
\graphicspath{{images/}}

\title{The GAPS Programme at TNG}
\subtitle{XXIV. An eccentric Neptune-mass planet near the inner edge of the BD-11 4672 habitable zone
            \thanks{Based on observations made with the Italian {\it Telescopio Nazionale Galileo} (TNG) operated by the {\it Fundaci\'on Galileo Galilei} (FGG) of the {\it Istituto Nazionale di Astrofisica} (INAF) at the {\it Observatorio del Roque de los Muchachos} (La Palma, Canary Islands, Spain). }
		    }

\author{ D.~Barbato             \inst{\ref{unito},\ref{oato},\ref{obsge}}
        \and M.~Pinamonti	    \inst{\ref{oato}}
        \and A.~Sozzetti	    \inst{\ref{oato}}
        \and K.~Biazzo		    \inst{\ref{oact},\ref{oarm}}
        \and S.~Benatti		    \inst{\ref{oapa}}
        \and M.~Damasso		    \inst{\ref{oato}}
        \and S.~Desidera	    \inst{\ref{oapd}}
        \and A.F.~Lanza		    \inst{\ref{oact}}
        \and J.~Maldonado	    \inst{\ref{oapa}}
        \and L.~Mancini		    \inst{\ref{unirm},\ref{mpia},\ref{oato}}
        \and G.~Scandariato	    \inst{\ref{oact}}
        \and L.~Affer		    \inst{\ref{oapa}}
        \and G.~Andreuzzi       \inst{\ref{fgg},\ref{oarm}}
        \and A.~Bignamini	    \inst{\ref{oats}}
        \and A.S.~Bonomo	    \inst{\ref{oato}}
        \and F.~Borsa		    \inst{\ref{oabrera}}
        \and I.~Carleo		    \inst{\ref{uniwes},\ref{oapd}}
        \and R.~Claudi		    \inst{\ref{oapd}}
        \and R.~Cosentino	    \inst{\ref{fgg}}
        \and E.~Covino		    \inst{\ref{oana}}
        \and A.F.M.~Fiorenzano  \inst{\ref{fgg}}
        \and P.~Giacobbe	    \inst{\ref{oato}}
        \and A.~Harutyunyan	    \inst{\ref{fgg}}
        \and C.~Knapic	        \inst{\ref{oats}}
        \and G.~Leto		    \inst{\ref{oact}}
        \and V.~Lorenzi         \inst{\ref{fgg},\ref{iac}}
        \and A.~Maggio		    \inst{\ref{oapa}}
        \and L.~Malavolta	    \inst{\ref{oact}}
        \and G.~Micela		    \inst{\ref{oapa}}
        \and E.~Molinari	    \inst{\ref{oaca}}
        \and M.~Molinaro	    \inst{\ref{oats}}
        \and V.~Nascimbeni	    \inst{\ref{oapd}}
        \and I.~Pagano		    \inst{\ref{oact}}
        \and M.~Pedani          \inst{\ref{fgg}}
        \and G.~Piotto		    \inst{\ref{unipd}}
        \and E.~Poretti		    \inst{\ref{fgg},\ref{oabrera}}
        \and M.~Rainer		    \inst{\ref{oafi}}
        }
\institute{ Dipartimento di Fisica, Universit\`{a} degli Studi di Torino, via Pietro Giuria 1, I-10125 Torino, Italy \label{unito}
            \and INAF – Osservatorio Astrofisico di Torino, Via Osservatorio 20, I-10025 Pino Torinese, Italy \label{oato}
            \and Observatoire de Gen\`{e}ve, Universit\'{e} de Gen\`{e}ve, 51 Chemin des Maillettes, CH-1290 Sauverny, Switzerland \label{obsge}
            \and INAF – Osservatorio Astrofisico di Catania, Via S. Sofia 78, I-95123, Catania, Italy \label{oact}
            \and INAF – Osservatorio Astronomico di Roma, Via Frascati 33, 00078 Monte Porzio Catone, Italy \label{oarm}
            \and INAF – Osservatorio Astronomico di Palermo, Piazza del Parlamento 1, I-90134, Palermo, Italy \label{oapa}
            \and INAF – Osservatorio Astronomico di Padova, Vicolo dell’Osservatorio 5, I-35122, Padova, Italy \label{oapd}
            \and Dipartimento di Fisica, Università di Roma Tor Vergata, Via della Ricerca Scientifica 1, 00133 Roma, Italy \label{unirm}
            \and Max Planck Institute for Astronomy, Königstuhl 17, 69117 Heidelberg, Germany \label{mpia}
            \and Fundación Galileo Galilei - INAF, Rambla José Ana Fernandez Pérez 7, E-38712 Breña Baja, TF, Spain \label{fgg}
            \and INAF – Osservatorio Astronomico di Trieste, via Tiepolo 11, I-34143 Trieste, Italy \label{oats}
            \and INAF – Osservatorio Astronomico di Brera, Via E. Bianchi 46, I-23807 Merate (LC), Italy \label{oabrera}
            \and Astronomy Department and Van Vleck Observatory, Wesleyan University, Middletown, CT 06459, USA \label{uniwes}
            \and INAF – Osservatorio Astronomico di Capodimonte, Salita Moiariello 16, I-80131, Napoli, Italy \label{oana}
            \and Instituto de Astrof\'{i}sica de Canarias, C/V\'{i}a L\'{a}ctea s/n, 38205 La Laguna, Spain \label{iac}
            \and INAF – Osservatorio Astronomico di Cagliari, Via della Scienza 5, I-09047 Selargius (CA) , Italy \label{oaca}
            \and Dipartimento di Fisica e Astronomia "G. Galilei", Universit\`{a} di Padova, Vicolo dell'Osservatorio 3, I-35122 Padova, Italy \label{unipd}
            \and INAF – Osservatorio Astrofisico di Arcetri, Largo E. Fermi 5, I-50125 Firenze, Italy \label{oafi}
            }
            
\date{Received <date> / Accepted <date>}
\abstract{With the growth of comparative exoplanetology, it is increasingly clear that investigating the relationships between inner and outer planets plays a key role in discriminating between competing formation and evolution models. To do so, it is important to probe the inner region of systems hosting long-period giants in search for undetected lower-mass planetary companions.}
{In this work we present our results on the K-dwarf star BD-11~4672, already known to host a long-period giant planet, as the first output of a subsample of the GAPS programme specifically aimed at assessing the impact of inefficient migration of planets formed beyond the snowline by searching for Neptune-mass and super-Earths planetary companions of known longer-period giants.}
{The high-precision HARPS-N observations of BD-11~4672 are used in conjunction with literature time series in order to search for additional inner planetary signals to be fitted using differential evolution Markov chain Monte Carlo. The long-term stability of the new orbital solutions is tested by N-body dynamical simulations.}
{We report the detection of BD-11~4672~c, a new Neptune-mass planet with an orbital period of $74.20_{-0.08}^{+0.06}\,\rm{d}$, eccentricity $0.40_{-0.15}^{+0.13}$, semimajor axis $0.30\pm0.01\,\rm{au}$ and minimum mass $15.37_{-2.81}^{+2.97}\,M_\oplus$ orbiting slightly outside the inner edge of the optimistic circumstellar habitable zone. In order to assess its impact on the dynamical stability of the habitable zone we compute the angular momentum deficit of the system, showing that planet c has a severe negative impact on the stability of possible additional lower-mass temperate planets. The BD-11~4672 system is notable for its architecture, hosting both a long-period giant planet and an inner lower-mass planet, the latter being also among the most eccentric Neptune-mass planets known at similar periods.}
{}

\begin{document}
    \maketitle
  
    \section{Introduction} \label{sec:introduction}
        \par The large number and variety of exoplanetary systems known so far has enabled the field of comparative exoplanetology to thrive, producing in recent years a number of studies focusing on planetary mass distributions and system architecture and their dependences on the characteristics of their host stars \citep[e.g.][]{winn2015,hobson2017,johnson2010,wittenmyer2020a}. As the competing formation and migration models of both terrestrial and giant planets produce different outcomes in terms of orbits and mass hierarchy \citep{raymond2008,cossou2014,schlaufman2014,morbidelli2016,izidoro2015a,izidoro2015b,bitsch2015,lambrechts2019}, the study of the architecture of known exoplanetary systems is a key factor in supporting or disproving the current theoretical models.
        \par In particular, the various formation and migration models of close-in super-Earths and Neptunes tend to predict different configurations of the inner regions of planetary systems in the presence of long-period giant planets. In the inward migration model, super-Earth embryos form in the outer protoplanetary disk and migrate inwards to become hot super-Earths or hot Neptunes; however if the innermost embryo grows into a gas giant planet it will block the inward migration of more distant cores \citep{izidoro2015a,izidoro2015b}. This model therefore results in a lack of inner lower-mass planets in the presence of outer giant planets. If cold Jupiters are instead formed at larger orbital distances \citep{bitsch2015} any subgiant planets formed near the snowline are able to migrate inward and become hot Neptunes and super-Earths, resulting in the presence of both close-in lower-mass bodies and long-period giant planets.
        \par Although the search for super-Earths or Neptune-mass planets on inner orbits in the presence of long-period giant planets is clearly fundamental in advancing our understanding of the dynamical history of planetary systems, technical and strategic issues have often severely impacted this endeavour. However, encouraging signs on our growing understanding in the characterization of the inner region of planetary systems are provided by recent observational programs and instrumentation well suited for such an endeavour and in the results of comparative exoplanetology. An analysis conducted in \cite{zhu2018} on inner super-Earths and Neptunes discovered by radial velocities and outer Jupiter analogs suggests that the majority of systems hosting cold Jupiters should also feature additional inner sub-Neptune planetary bodies. More recently, the 18-years of observations collected by the Anglo-Australian Planet Search reported in \cite{wittenmyer2020b} suggests that the occurrence rate of giant planets orbiting Solar-type stars having semimajor axis $a>1\,\rm{au}$ is $\sim7$\%, about eight times higher than those on closer-in orbits. Such a predominance of long-period giant planet provides fertile grounds for the search of additional lower-mass planets in the inner orbital region and the study of early formation and inward migration mechanisms. In a previous survey \citep{barbato2018}, we monitored 20 Sun-like stars known to host long-period giant planets, and derived an occurrence rate of inner lower-mass planets $< 10 \%$ for $M\sin{i}>10\,M_\oplus$. However, the final completeness of our survey was negatively impacted by the average number of datapoint collected per star being lower than initially foreseen, severely limiting the sensitivity to lower-mass companions. Indeed, the real occurrence rate of such systems may actually be higher, as additional inner low-mass planets in the presence of outer giant planets could be detected with intensive sampling and longer observational timespans \citep[e.g.][]{benatti2020}.
        \par In this context, K and M dwarf stars represent a fertile field for the search of inner super-Earths and Neptune-mass planetary companion. Such stars represent the most numerous stellar population in the Milky Way \citep[e.g.][]{chabrier2000,winters2015} and their low mass, size and temperature make them prime candidates for the search of planetary systems, easing the detection of signals produced by sub-Neptune planetary companions and hosting a variety of systems featuring super-Earths and Neptune-mass planets \citep{bonfils2012,dressing2013,crossfield2015,astudillodefru2017b,pinamonti2018}. While it has been observed that dwarfs less massive than our Sun are less likely to host a giant planet \citep{butler2004,johnson2007,johnson2010,bonfils2013} as expected by theoretical studies \citep[e.g.][]{laughlin2004,mordasini2009a,mordasini2009b}, the discovery of M-dwarfs hosting planetary systems featuring both giant planets and sub-Neptunians companions, such as the GJ~676~A system \citep{forveille2011,angladaescude2012b} hosting two inner sub-Neptunian planets and two outer giant planets, suggests that the formation and survival of multiple planetary systems featuring a wide range of planetary masses is not completely suppressed around dwarf stars.
        \par The observational programme Global Architecture of Planetary Systems (GAPS, see \citealt{covino2013,desidera2013}) aims at investigating the variety and origins of the architecture of exoplanetary systems with the High Accuracy Radial velocity Planet Searcher in the Northern hemisphere (HARPS-N, \citealt{cosentino2012}) at the Telescopio Nazionale \textit{Galileo} (TNG) in La Palma. Within this context, a subset of planet-hosting K and M dwarfs has been selected and observed during the last years in order to search for inner super-Earths and Neptune-mass planetary companions to outer long-period giants and assessing the impact of inefficient migration of planets formed beyond the snowline of such dwarf stellar hosts.
        \par In this paper we present the first result of the GAPS observations on this stellar sample, namely the new characterization of the planetary system orbiting the K-dwarf star BD-11~4672 (also known as GJ 717). \cite{moutou2011} first reported hints of the presence of a planetary companion around BD-11~4672 following observations with the High Accuracy Radial velocity Planet Searcher (HARPS, \citealt{mayor2003}) at the ESO La Silla 3.6m telescope. Additional HARPS monitoring \citep{moutou2015} led to the detection of the planetary companion BD-11~4672~b, with a period of $1667_{-31}^{+33}$ days, and  minimum mass $M\sin{i}=0.53\pm0.05\,M_{\rm{Jup}}$.
        \par In Sect. \ref{sec:star-parameters} we provide updates on the parameters of the stellar host. In Sect. \ref{sec:time series} we provide an overview of the HARPS-N observations, analyse the stellar activity cycles, and fit our HARPS-N radial velocity measurements in conjunction with archival HARPS data. In Sect. \ref{sec:dynamics} we assess the dynamical stability of the system, focusing on the circumstellar habitable zone, before concluding and discussing the results of this work in Sect. \ref{sec:conclusions}.
        
    \section{Stellar parameters}    \label{sec:star-parameters}
        \begin{table}
          \caption{Elemental abundances for star BD-11~4672. The first errors refer to the measure of the equivalent width, while the errors in parentheses are obtained from the 
                    root sum square of the abundance error caused by uncertainties on $T_{\rm eff }$, $\log{g,}$ and $\xi$.}       \label{table:star-abundances}
            \centering
            \begin{tabular}{l c}
              \hline\hline
                \multicolumn{2}{c}{Elemental abundances (dex)}\\
              \hline
                [\ion{Fe}{i}/H]                 & $-0.35\pm0.15$ ($+-0.06$) \\[3pt]
                [\ion{Fe}{ii}/H]                & $-0.35\pm0.27$ ($+-0.24$) \\[3pt]
                [\ion{Na}{i}/H]                 & $-0.37\pm0.03$ ($+-0.14$) \\[3pt]
                [\ion{Mg}{i}/H]                 & $-0.48\pm0.19$ ($+-0.02$) \\[3pt]
                [\ion{Al}{i}/H]                 & $-0.24\pm0.14$ ($+-0.08$) \\[3pt]
                [\ion{Si}{i}/H]                 & $-0.39\pm0.16$ ($+-0.09$) \\[3pt]
                [\ion{Ca}{i}/H]                 & $-0.17\pm0.06$ ($+-0.19$) \\[3pt]
                [\ion{Ti}{i}/H]                 & $-0.18\pm0.08$ ($+-0.21$) \\[3pt]
                [\ion{Ti}{ii}/H]                & $-0.61\pm0.06$ ($+-0.09$) \\[3pt]
                [\ion{Cr}{i}/H]                 & $-0.27\pm0.11$ ($+-0.13$) \\[3pt]
                [\ion{Cr}{ii}/H]                & $-0.30\pm0.08$ ($+-0.12$) \\[3pt]
                [\ion{Ni}{i}/H]                 & $-0.49\pm0.18$ ($+-0.04$) \\[3pt]
                [\ion{Ba}{ii}/H]                & $-0.43\pm0.07$ ($+-0.19$) \\[3pt]
                [\ion{C}{i}/H]                  & $+0.24\pm0.03$ ($+-0.16$) \\[3pt]
                [\ion{Y}{ii}/H]                 & $-0.62\pm0.16$ ($+-0.09$) \\[3pt]
                [\ion{Eu}{ii}/H]                & $+0.04\pm0.04$ ($+-0.07$) \\[3pt]
                [\ion{Nd}{ii}/H]                & $-0.44\pm0.38$ ($+-0.09$) \\[3pt]
                [\ion{Nd}{ii}/H]                & $+0.43\pm0.37$ ($+-0.13$) \\[3pt]
                [\ion{Zn}{i}/H]                 & $+0.11\pm0.04$ ($+-0.12$) \\[3pt]
              \hline
            \end{tabular}
        \end{table}
        \begin{table}
          \caption{Stellar parameters for BD-11~4672.}       \label{table:star-parameters}
            \centering
            \begin{tabular}{l c}
              \hline\hline
                \multicolumn{2}{c}{BD-11~4672 (GJ 717)}\\
              \hline
                $\alpha$ (J2000)\tablefootmark{a} 		                    & $18^h33^m28.8^s$\\[3pt]
                $\delta$ (J2000)\tablefootmark{a} 	        	            & $-11^{\degr}38^\prime9^{\prime\prime}$\\[3pt]
                $\pi$ (mas)\tablefootmark{a} 		        	            & $36.781\pm0.046$\\[3pt]
                $\mu_\alpha$ ($\rm{mas}\,\rm{yr}^{-1}$)\tablefootmark{a} 	& $-288.590\pm0.097$\\[3pt]
                $\mu_\delta$ ($\rm{mas}\,\rm{yr}^{-1}$)\tablefootmark{a} 	& $-235.560\pm0.089$\\[3pt]
                $B$ (mag)\tablefootmark{b} 			                        & $11.21\pm0.10$\\[3pt]
        		$V$ (mag)\tablefootmark{b} 			                        & $9.99\pm0.05$\\[3pt]
        		$R$ (mag)\tablefootmark{c} 		        	                & $9.60\pm0.10$\\[3pt]
                $G$ (mag)\tablefootmark{a}  	                            & $9.4851\pm0.0005$\\[3pt]
        		$J$ (mag)\tablefootmark{d} 			                        & $7.651\pm0.020$\\[3pt]
        		$H$ (mag)\tablefootmark{d}      		    	            & $7.031\pm0.030$\\[3pt]
        		$K$ (mag)\tablefootmark{d} 	        		                & $6.867\pm0.020$\\[3pt]
                Spectral Type\tablefootmark{e}	        	                & K7V\\[3pt]
                $M_\star$ ($M_\odot)$\tablefootmark{f}        	            & $0.651^{+0.031}_{-0.029}$\\[3pt]
                $R_\star$ ($R_\odot)$\tablefootmark{f} 	                    & $0.639^{+0.020}_{-0.022}$\\[3pt]
                $\rho_\star$ ($\rm{g}\,\rm{cm}^{-3}$)\tablefootmark{f}  	& $3.53^{+0.34}_{-0.28}$\\[3pt]
                $T_{\rm eff }$ (K)\tablefootmark{f} 			            & $4550\pm110$\\[3pt]
                $L_\star$ ($L_\odot$)\tablefootmark{f}                      & $0.157^{+0.019}_{-0.017}$\\[3pt]
                $\log{g}$ (cgs)\tablefootmark{f} 	        	            & $4.642^{+0.027}_{-0.025}$\\[3pt]
                Age (Gyr)\tablefootmark{f} 		            	            & $7.4^{+4.5}_{-4.9}$\\[3pt]
                $\log{R^{\prime}_{\rm{HK}}}$\tablefootmark{g}               & $-4.65$ \\[3pt]
                $P_{\rm rot}$ (d)\tablefootmark{g}                          & $\sim25$\\[3pt]
              \hline
            \end{tabular}
            \tablefoot{
                          \tablefoottext{a}{ retrieved from Gaia Data Release 2 \citep{gaia2018} }
                          \tablefoottext{b}{ retrieved from \cite{hog2000} }
                          \tablefoottext{c}{ retrieved from \cite{zacharias2012}}
                          \tablefoottext{d}{ retrieved from \cite{cutri2003} }
                          \tablefoottext{e}{ retrieved from \cite{skiff2014} }
                          \tablefoottext{f}{ obtained from the SED fitting discussed in Sect. \ref{sec:star-parameters} }
                          \tablefoottext{g}{ expected value obtained from the activity analysis discussed in Sect. \ref{subsec:activity} }
                        }
        \end{table}
        \begin{figure}
          \includegraphics[width=\linewidth]{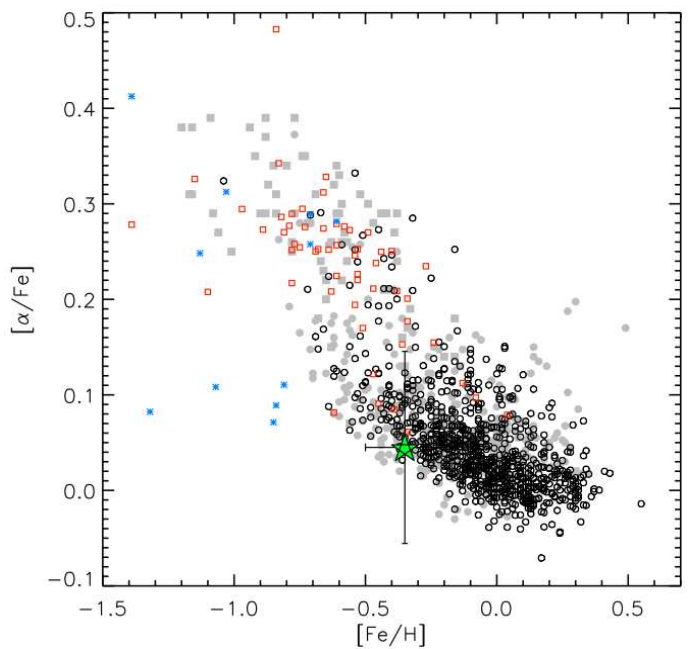}
          \caption{[$\alpha$/Fe] vs. [Fe/H] for BD-11~4672, here identified by the green star symbol. Thin-disc, thick-disc, and halo stars are shown with circles, squares, and asterisks, respectively 
                    (filled symbols: \citealt{soubiran2005}; open symbols: \citealt{adibekyan2012b}).
                  }
          \label{fig:star-population}
        \end{figure}
        \par In this section we present a new spectroscopic characterization of astrophysical properties of the host star BD-11~4672. We measured effective temperature $T_{\rm eff}$, surface gravity $\log{g}$, microturbulence velocity $\xi$, and iron abundance [Fe/H] through a method based on line equivalent widths. The measurement of equivalent widths was done through the software \texttt{IRAF} \citep{tody1993} from the coadded spectrum built from individual HARPS-N spectra used for the radial velocity measurements and described in Sect. \ref{sec:time series}. We then adopted the \cite{castelli2004} grid of model atmospheres and the spectral analysis package MOOG (\citealt{sneden1973}; 2017 version). $T_{\rm eff}$ was derived by imposing that the abundance of \ion{Fe}{i} is not dependent on the line excitation potentials, $\xi$ by obtaining the independence between \ion{Fe}{i} abundance and equivalent widths, and $\log{g}$ by the \ion{Fe}{i}/\ion{Fe}{ii} ionization equilibrium condition.
        \par With this method, we obtain the following stellar parameters: $T_{\rm eff}=4500\pm130\,\rm{K}$, $\log{g}=4.5\pm0.2$, $\xi=0.85\pm0.40\,\rm{km}\,\rm{s}^{-1}$, and $\rm{[Fe/H]}=-0.35\pm0.15\,\rm{dex}$. From these values of stellar parameters we followed the method detailed in \cite{damasso2015} and \cite{biazzo2015}, and references therein, to measure other elemental abundances differentially with respect to the Sun (see Table \ref{table:star-abundances}). The first value of the uncertainty on elemental abundance is obtained from the equivalent widths measurements, while the second value is the root sum square of the errors due to the uncertainties in stellar parameters. With an iron abundance of $\rm{[Fe/H]}=-0.35\pm0.15\,\rm{dex}$, we can consider the star BD-11~4672 as a metal-poor host star, although not as lacking in metal content as other planet-hosting stars we observed with HARPS-N within the GAPS programme \citep{barbato2019}.
        \par We also derived the projected rotational velocity $v\sin{i}$ through spectral synthesis of two regions around 6200 and 6700\,\AA, as done in \citet{barbato2019}. Using the same grid of model atmospheres and code and assuming a macroturbolence $v_{\rm macro}=2.0\,\rm{km}\,\rm{s}^{-1}$ \citep{valenti2005}, we find $v\sin{i}=1.0\pm0.5\,\rm{km}\,\rm{s}^{-1}$, which is below the HARPS-N spectral resolution, thus suggesting a slow stellar rotation, unless the star is observed nearly pole-on. \citet{moutou2015} did not measure the projected rotation velocity, and was only able to put an upper limit $v\sin{i} < 2\,\rm{km}\,\rm{s}^{-1}$, which is consistent with our findings.
        \par Considering the elemental abundances of the field stars listed in the catalogues by \cite{soubiran2005} and \cite{adibekyan2012b}, and applying the prescriptions reported in \cite{biazzo2015}, we can also investigate  which stellar population BD-11~4672 belongs to. We thus considered the abundances of the $\alpha$-elements Mg, Si, Ca, and Ti. Fig. \ref{fig:star-population} shows the position of BD-11~4672 in the [$\alpha$/Fe] versus [Fe/H] diagram. Based on these chemical indicators, the star seems to be a member of the thin-disc population. Additionally, we can classify BD-11~4672 on the basis of its Galactic kinematics, using Gaia Data Release 2 proper motions \citep{gaia2018} to compute the thick-to-thin disc probability ratio $TD/D$ defined in \cite{bensby2003}. For BD-11~4672 we find $TD/D=0.45$; while this value hints at an intermediate kinematics between the two populations, the tests on different population normalizations conducted in \cite{bensby2005} show that stars with $TD/D<0.6$ can be usually safely regarded as thin disc stars. Both chemical composition and kinematics criteria therefore point towards BD-11~4672 as being a member of the thin-disc stellar population.    
    	\par In order to provide updated estimates of all stellar parameters we fit the Spectral Energy Distribution (SED) via the MESA Isochrones and Stellar Tracks (MIST) \citep{dotter2016,choi2016} through the \texttt{EXOFASTv2} suite \citep{eastman2019}. We fit the available archival magnitudes listed in Table \ref{table:star-parameters}, imposing gaussian priors on $T_{\rm eff }$ and [Fe/H] as obtained by the ionization equilibrium and on parallax $\pi$ based on the Gaia DR2 astrometric measurement, after correcting the DR2 value for the systematic offset of $-82\pm33\,\mu\rm{as}$ as reported in \citep{stassun2018}; this astrometric prior helps constraining the stellar radius and improves the precision of the stellar parameters resulting from the SED fitting procedure. The resulting stellar parameters, listed in Table \ref{table:star-parameters}, are generally comparable within 2$\sigma$ to the stellar characterization reported in \cite{moutou2015}, with the exception of the star's mass and radius, for which we find larger values differing from the \cite{moutou2015} ones by 2.5$\sigma$ and 4.3$\sigma$ respectively.
    	\par With these updated stellar parameters, we use the 1-D radiative-convective, cloud-free climate model detailed in \cite{kopparapu2013,kopparapu2014} to compute the inner and outer boundaries of the circumstellar habitable zone. Considering the runaway and maximum greenhouse limits, we characterize the conservative habitable zone as spanning from 0.42 to 0.77 au. Using instead the recent Venus and early Mars limits, we find the optimistic habitable zone ranging from 0.33 to 0.81 au.
    	
    \section{Spectroscopic time series and analyses}  \label{sec:time series}
            \begin{figure}
                \centering
                \includegraphics[width=\linewidth]{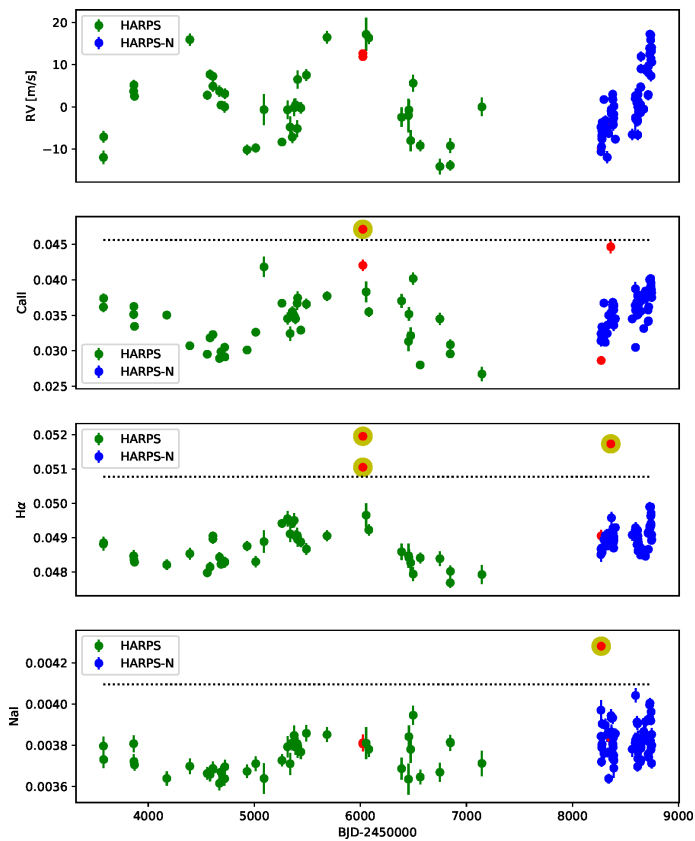}
                \caption{Combined HARPS and HARPS-N time series of the RVs and activity indices of BD-11~4672. The red dots represent the outliers identified in the activity time series, and the yellow shades mark the outliers identified in each time series. The dotted black lines indicates the threshold value of three standard deviations from the median.}
                \label{fig:BD-114672-act-time-series}
            \end{figure}
    	\par We observed BD-11~4672 with HARPS-N from May 2018 ($\rm{BJD}=2458268.6$) to September 2019 ($\rm{BJD}=2458747.4$), obtaining a total of 68 datapoints over a timespan of 479 days with a mean exposure time of 1100 s and a mean signal-to-noise ratio (S/N) of 66. The spectra collected were reduced using the HARPS-N Data Reduction Software (DRS, \citealt{lovis2007}), while the radial velocities extraction was performed with the Template-Enhanced Radial velocity Re-analysis Application pipeline (TERRA, \citealt{angladaescude2012a}), which proved to provide more precise radial velocity measurements for late-type stars. The thusly obtained radial velocity time series has a median radial velocity uncertainty of 0.95 $\rm{m}\,\rm{s}^{-1}$ and a weighted root mean square (w.r.m.s.) of 7.23 $\rm{m}\,\rm{s}^{-1}$.
    	\par In the following analysis we use our HARPS-N measurements in combination with the HARPS observations presented in \cite{moutou2015} and three additional HARPS archival datapoints collected from July 2014 to early May 2015, all of which we newly reduced using TERRA. While an additional archival HARPS datapoint at epoch 24557851.82 is available, it was collected after the late May 2015 optical fibre upgrade \citep{locurto2015}, therefore introducing an instrumental offset depending on the stellar spectral type \citep{trifonov2020}. Rather than treating this single datapoint as independent from the rest of the time series, we exclude this point from all analysis. The historical time series is then composed by 43 datapoints over a timespan of 3568 days with median error of 1.38 $\rm{m}\,\rm{s}^{-1}$ and w.r.m.s. of 8.55 $\rm{m}\,\rm{s}^{-1}$.
    	\par Therefore, the full HARPS and HARPS-N time series we take into exam is composed of a total of 111 radial velocity measurements over 5170 days with a median uncertainty of 1.14 $\rm{m}\,\rm{s}^{-1}$ and w.r.m.s. of 7.77 $\rm{m}\,\rm{s}^{-1}$. The complete lists of HARPS and HARPS-N radial velocity measurements, in addition to the time series of the stellar activity indicators that will be discussed in Sect. \ref{subsec:activity}, are shown in Tables \ref{table:rv-harps} and \ref{table:rv-harpsn}.
        \par In order to study the stellar activity as described in Sect. \ref{subsec:activity}, we computed a series of activity indices for all the HARPS and HARPS-N spectra at our disposal, to obtain consistent time series of indicators in order to better analyse the long-term evolution of the stellar activity. Following the procedure from \citet{gomesdasilva11} we derived the activity indices for the Ca~{\sc ii}  H\&K, H$\alpha$, and Na~{\sc i} D$_{\rm 1}$ D$_{\rm 2}$ stellar spectral lines.
    
        \subsection{Activity} \label{subsec:activity}
            \begin{figure*}
                \centering
                \includegraphics[width=\linewidth]{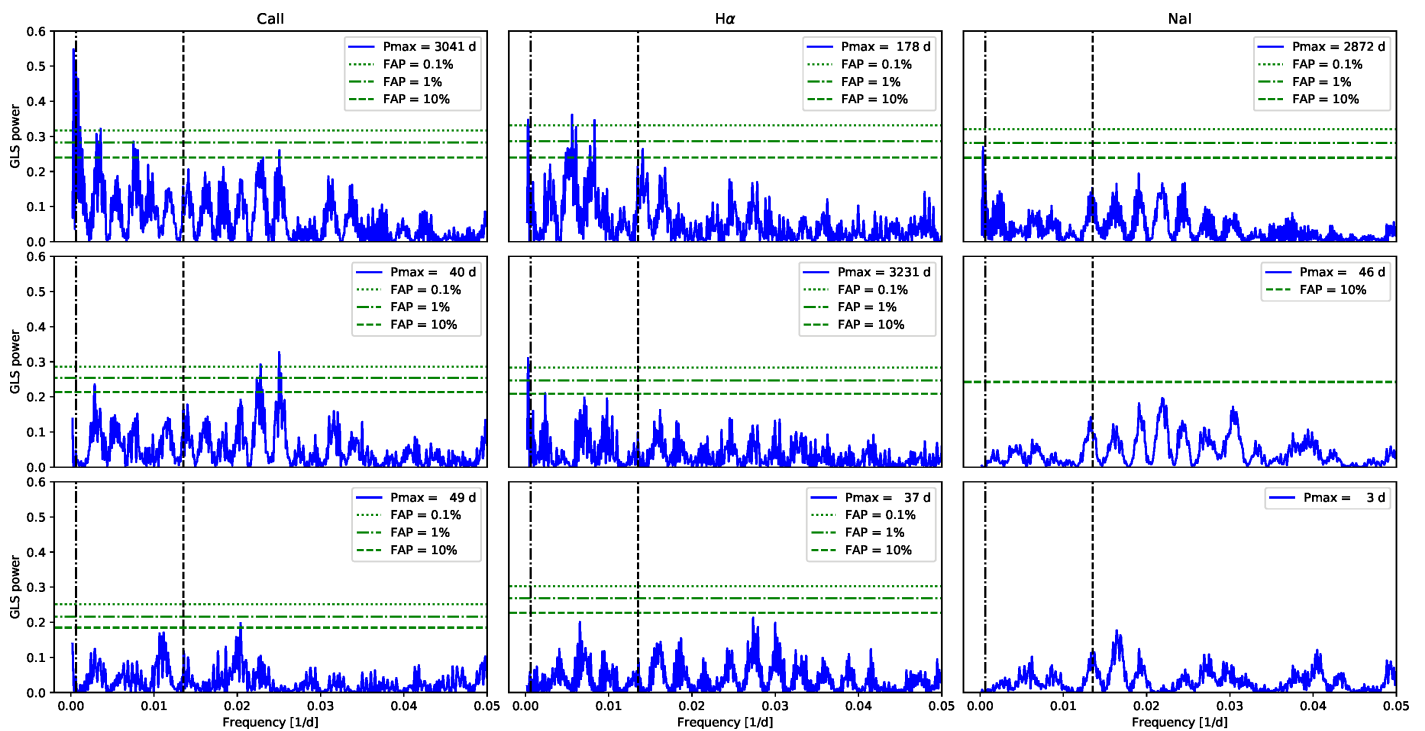}
                \caption{Generalized Lomb-Scargle periodograms of the activity indices time series (top row) and residuals after pre-whitening (middle and bottom row). The period value of the principal peak of each periodogram is noted in the legend as $P\rm{max}$, and the dashed and dot-dashed black vertical lines indicates the signals discussed in Sect. \ref{subsec:time series-fit}. The horizontal green lines indicates the False Alarm Probability (FAP) levels, as indicated in the legend. The FAP levels are computed via 10000-iterations bootstrap. In the residuals periodograms only the FAP levels less significant than the principal peak of the previous periodograms are marked.}
                \label{fig:activity-periodograms}
            \end{figure*}
            We studied the stellar activity of BD-114672, expanding the analyses from \citet{moutou2011} and \citet{moutou2015}. They searched for correlations between the RVs and activity indices derived from the HARPS spectra, and for signals of activity evolution on time scales comparable to the orbital period of BD-11~4672~b. Even if the results from \citet{moutou2011} were inconclusive, \citet{moutou2015} found no evidence of the 1667 d signal to be produced by a magnetic cycle.
            \par First we checked for the presence of possible outliers in the activity indices time series, which can be caused by stellar flares or other transient phenomena. To do so we identified the epochs for which any activity index deviated more than three standard deviations from the median of the respective time series. These observational epochs were discarded from all the activity, asymmetry and RV time series in all the following analyses, since the data could be affected by transient phenomena which are not considered by our models.
            \par In the bottom three panels of Fig. \ref{fig:BD-114672-act-time-series} we show the time series for the activity indices, highlighting the four outliers identified with the aforementioned procedure (BJD$=$2456023.85, 2456023.89, 2458270.69 and 2458361.38). There clearly appears to be some long-term modulation in the  Ca~{\sc ii}  H\&K, and H$\alpha$ time series and, if less evident, in the Na~{\sc i} D$_{\rm 1}$ D$_{\rm 2}$. To study these potential signals we computed the Generalized Lomb Scargle periodogram \citep[GLS,][]{zechmeisterkurster2009} of the activity time series. The results are shown in the top row of Fig. \ref{fig:activity-periodograms}. The Ca~{\sc ii}  H\&K and Na~{\sc i} D$_{\rm 1}$ D$_{\rm 2}$ periodograms are dominated by long-period signals, with similar periods of $P\simeq 3000\,\rm{d}$, while the H$\alpha$ shows a significant peak around that period, but is dominated by a shorter periodicity.
            \par
\begin{figure}
                \centering
                \includegraphics[width=\linewidth]{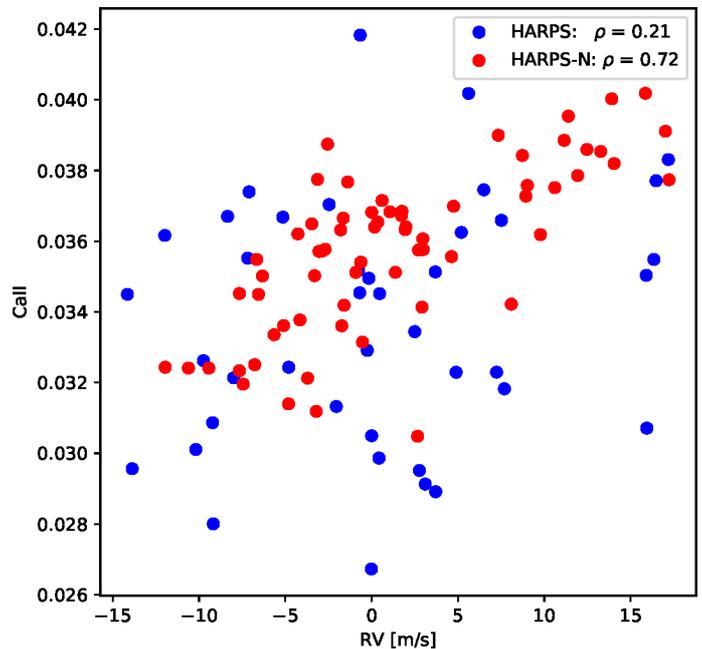}
                \caption{Ca~{\sc ii}  H\&K index time series as a function of the radial velocity measurements.}
                \label{fig:rv-caii-corr}
            \end{figure}

 Since also the RV time series, shown in the top panel of Fig. \ref{fig:BD-114672-act-time-series}, presents a long-term periodicity, corresponding to the orbital period $P_b \simeq 1700$ d of BD-114672 b \citep{moutou2015}, we studied the correlation between the RV and activity indices time series. We computed the Pearson correlation coefficients and found no significant correlation ($\left | \rho \right | < 0.3$) for any indices except Ca~{\sc ii}  H\&K, which showed a moderate correlation $\rho_\text{Ca~{\sc ii}} = 0.45$. As can be seen in Fig. \ref{fig:rv-caii-corr}, this correlation is dominated by the HARPS-N data, which cover only small fractions of the 3000 d periodicity dominating the activity time series and of the 1700 d orbital signal of BD-114672 b, in which both the signals show an ascending trend: computing the Pearson correlation coefficient on the HARPS time series, which has a timespan longer than both the involved periods, the correlation drops to $\rho_\text{Ca~{\sc ii}} = 0.21$. This suggests that the moderate value of $\rho_\text{Ca~{\sc ii}}$ for the combined time series is only an artefact of the uneven sampling of the periodic signals dominating the RV and activity data. Moreover, while performing the fits described in Sect. \ref{subsec:time series-fit}, we repeated the computation of the Pearson correlation coefficients with the activity indices on the RV residuals of the fitted models: no increase in the correlation was observed, thus confirming that the emerging signals shown in Fig. \ref{fig:rv-gls} are not activity-related.
            \par Additional significant peaks can be seen in the top row of Fig. \ref{fig:activity-periodograms}, and thus we performed a residuals analysis of the activity time series after the subtraction of the dominant peak in each GLS periodogram. The results are shown in the middle and bottom rows of Fig. \ref{fig:activity-periodograms}, for the analyses of the first and second residuals respectively. We can see in the middle row that the Ca~{\sc ii}  H\&K and Na~{\sc i} D$_{\rm 1}$ D$_{\rm 2}$ residuals both show signals around $\sim 40\,\rm{d}$, even if in the case of Na~{\sc i} D$_{\rm 1}$ D$_{\rm 2}$ the peak is not significant; in the H$\alpha$ residuals the 3000 d long period signals becomes dominant. Only the second residuals of the Ca~{\sc ii}  H\&K show a significant signal, which is again close to 40 d.
            We thus have three dominant periodicities in the analysed activity indices: $\sim 3000$, $\sim 200$, and $\sim 40\,\rm{d}$. The longest periodicity, which is present in all three activity indicators, can be attributed to a solar-like activity cycle, since similar cycles spanning around 7-10 yr have been observed on other stars of similar spectral type \citep{borosaikiaetal2018}.
            \par We collected the available All Sky Automated Survey \citep[ASAS,][]{pojmanski1997} photometric timeseries for BD-11~4672, composed of 650 datapoints spanning over 3173 days from February 2001 to November 2009. We performed a GLS analysis of these data, and identified a dominant periodicity of $\sim3000\,\rm{d}$, consistent with the long-period activity cycle we identify in the HARPS and HARPS-N activity indices. However, the relatively short timespan of the available ASAS photometry does not allow for a more precise characterization of this long-term periodicity.
            \par Regarding the shorter periodicities, \cite{scandariato2017} notes that in cold and relatively inactive stars periods around $\sim200\,\rm{d}$ are usually linked to the evolution timescale of active regions, and for the same stars rotation periods are typically around $\sim30-40\,\rm{d}$.
            No information is present in the literature on the rotation period of BD-11~4672, which is usually linked to the dominant signals in the activity indicators. From the value of $v\sin{i} $ derived in Sect. \ref{sec:star-parameters}, we can estimate the maximum rotation period $P_{\rm{rot, max}} = 30\pm 15\,\rm{d}$, even if the large errorbars on the measurements of $v\sin{i} $ translate into large uncertainties on this estimates. Moreover, BD-11~4672 has not been observed by the Transiting Exoplanet Survey Satellite \citep[TESS,][]{ricker2015} during its nominal mission. However, as the TESS extended mission is set to observe also in the ecliptic, it may help obtaining a photometric measurement of the rotation period of BD-11~4672 in the near future.
            \par A common technique to estimate the value of the stellar rotation period is from the empirical relationships between stellar rotation and the mean activity level \citep[e.g.][]{noyesetal1984,mamajekhillenbrand2008}. The mean activity level is usually measured by means of the $\log{R'_{\rm HK}}$, which was defined for FGK-type stars by \citet{noyesetal1984}, but is not calibrated for lower-mass stars, as late-K and M dwarfs. Different calibration procedures have been proposed in recent years to compute the $\log{R'_{\rm HK}}$ for late-type stars, such as those by \citet{astudillodefru2017a} and \citet{suarezmascarenoetal2018}. We used these two calibration procedures to derive the mean values of $\log{R'_{\rm HK}}$ from the S-index computed from our HARPS-N spectra, and obtained the similar values of $\log{R'_{\rm HK}} = -4.65 $ and $\log{R'_{\rm HK}} = -4.71 $ from \citet{astudillodefru2017a} and \citet{suarezmascarenoetal2018} respectively. These authors also provide activity-rotation relationships, from which we obtained estimates for BD-11~4672's rotation period of  $P_{\rm rot} = 24.7 \pm 2.5$ d and $P_{\rm rot} = 19.6 \pm 2$ d, respectively. The uncertainties are estimated from the dispersion of the activity-rotation relationship, which is on the order of $\sim 10 \%$ \citep{astudillodefru2017a}, and show that the two values are compatible within $2 \sigma$.
    
        \subsection{Orbital fitting of the radial velocities} \label{subsec:time series-fit}
            \begin{figure}
                \centering
                \includegraphics[width=\linewidth]{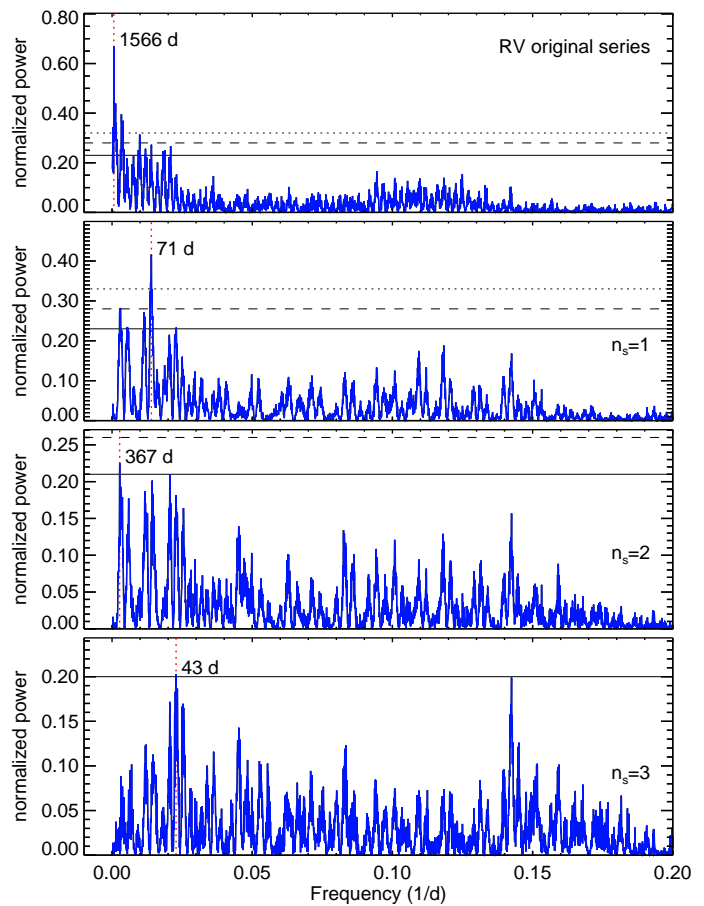}
                \caption{Generalized Lomb-Scargle periodogram of the HARPS and HARPS-N data of BD-11~4672 (top panel) and the residuals obtained after removing the $n_s-$signals best fit solution (second to fourth panel). In each panel, the periodogram's most significant frequency is identified by a vertical dotted red line and labeled by its corresponding period, while the horizontal solid, dashed and dotted lines respectively identify the 10\%, 1\% and 0.1\% FAP levels.}
                \label{fig:rv-gls}
            \end{figure}
            After excluding from the radial velocity time series the four outliers discussed in Sect. \ref{subsec:activity}, we have a total of 107 datapoints, 41 of which being the archival HARPS measurements and 66 from our HARPS-N monitoring of the star.
            \begin{figure}
                \centering
                \includegraphics[width=\linewidth]{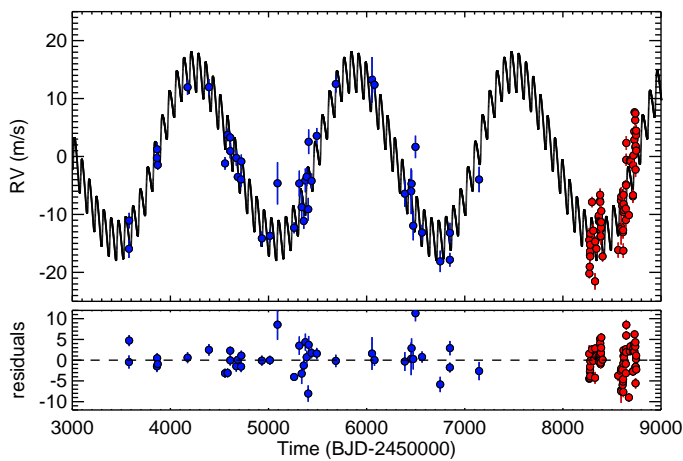}
                \caption{Orbital fit for the two-planet solution of system BD-11~4672. In the top panel the best fit solution is shown as a black curve over the literature datapoints from HARPS (blue) and our HARPS-N observations (red). The bottom panel shows the residual radial velocities.}
                \label{fig:orbital-fit-2pl}
            \end{figure}
            \begin{figure}
                \centering
                \includegraphics[width=\linewidth]{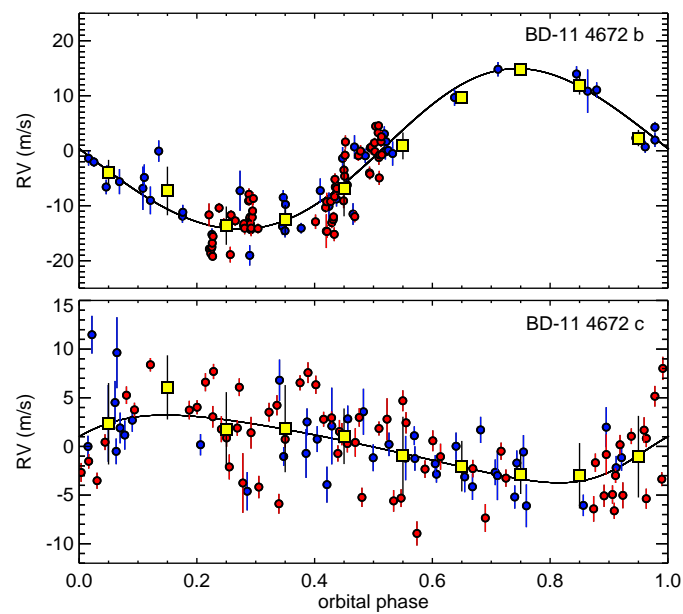}
                \caption{Phase-folded radial velocity curves for our two-planets solution of the BD-11~4672 system. The best fit solution for each planet is shown as a black curve over the literature datapoints from HARPS (blue) and our HARPS-N observations (red). The yellow squares and their error bars represent the binned averages and standard deviations of the measurements.}
                \label{fig:phase-fit-2pl}
            \end{figure}
            \par We started by investigating the main peak of the after pre-whitening full time series GLS periodogram (see top panel of Fig. \ref{fig:rv-gls}), located at 1566 d with a false alarm probability (FAP) of 0.01\% as calculated via bootstrap method, in order to provide an updated orbital solution for the known planet b first reported in \cite{moutou2015}. We searched for the best fit orbital solution through a differential evolution Markov chain Monte Carlo method based on the \texttt{EXOFAST} suite \citep{eastman2013,desidera2004}. For a single-Keplerian, eccentric solution the nine free parameters were inferior conjunction epoch $T_0$, orbital period $P$, $\sqrt{e}\cos{\omega}$, $\sqrt{e}\sin{\omega}$, semi-amplitude $K$, and, for each instrument considered (HARPS and HARPS-N), a zero-point radial velocity $\gamma$ and an uncorrelated stellar jitter term $j$ added in quadrature to the radial velocity measurement error. Uninformative priors were used for all parameters, explicitly setting minimum and maximum boundaries only for $P$ and $T_0$ in order to better explore the parameter region most relevant to each analysed signal. Eighteen chains were run simultaneously and reached convergence and good mixing according to the criteria established in \cite{eastman2013}.
            \par We find a single-Keplerian solution (see second column of Table \ref{table:planets-parameters}) with semiamplitude $K_{\rm b}=14.58_{-1.04}^{+1.08}$ $\rm{m}\,\rm{s}^{-1}$, period $P_{\rm b}=1628\pm12\,\rm{d}$ and eccentricity $e_{\rm b}=0.10\pm0.06$, from which we derive minimum mass $M_{\rm b}\sin{i}=0.63\pm0.05\,M_{\rm{Jup}}$ and semimajor axis $a_{\rm b}=2.35\pm0.04\,\rm{au}$. We note that the Keplerian parameters of this solution are all comparable to those reported in \cite{moutou2015} within 1$\sigma$. Due to the larger estimate of the stellar mass derived in Sect. \ref{sec:star-parameters}, we obtain larger values of minimum mass and semi-major axis. Also, we note that the $e_{\rm b}$ values of both solutions are compatible with zero within 2$\sigma$.
            \par The periodogram of the residuals to the single-planet solution (shown in the second panel of Fig. \ref{fig:rv-gls}) features a significant peak at 71 d with $\rm{FAP}=0.01\%$ lacking any counterpart in the activity indices discussed in Sect. \ref{subsec:activity} and shown in Fig. \ref{fig:activity-periodograms}, suggesting the presence of an additional planetary companion.
            \par We find the best fit two-Keplerian solution (see third column of Table \ref{table:planets-parameters}) to have orbital periods $P_{\rm b}=1631\pm13\,\rm{d}$ and $P_{\rm c}=74.18_{-0.12}^{+0.10}\,\rm{d}$, semiamplitudes $K_{\rm b}=14.59_{-0.98}^{+1.01}$ $\rm{m}\,\rm{s}^{-1}$ and $K_{\rm c}=3.48_{-0.67}^{+0.78}$ $\rm{m}\,\rm{s}^{-1}$ and eccentricities $e_{\rm b}=0.08_{-0.05}^{+0.06}$ and $e_{\rm c}=0.34_{-0.17}^{+0.18}$; this solution implies for planet b minimum mass $M_{\rm b}\sin{i}=0.63\pm0.05\,M_{\rm{Jup}}$ and semimajor axis $a_{\rm b}=2.36\pm0.04\,\rm{au}$. Again, we note that $e_{\rm b}$ is compatible with zero within 2$\sigma$.
            In order to compare this two-planet solution with the single-planet one previously found, we compute the Bayesian Information Criterion (BIC) values of each solution as:
            \begin{equation}
             \rm{BIC}=k\log{n}-2\log{\mathscr{L}}
            \end{equation}
            being $k$ the number of model parameters, $n$ the datapoint number and $\log\mathscr{L}$ the maximum log-likelihood computed by \texttt{EXOFAST}. We find this two-Keplerian solution to describe the time series significantly better than the single-planet one, having respectively BIC$_{\rm 2pl}=425.78$ and BIC$_{\rm 1pl}=440.83$, the difference of $\Delta\rm{BIC}=-15.05$ between the two solutions representing a strong evidence in favour of the two-Keplerian solution.
            \par We therefore report the detection of a second planetary companion orbiting at 74 days around star BD-11~4672 and for which we derive minimum mass $M_{\rm c}\sin{i}=16.59_{-2.97}^{+3.14}\,M_\oplus$ and semimajor axis $a_{\rm c}=0.30\pm0.01\,\rm{au}$; in Fig. \ref{fig:orbital-fit-2pl} and \ref{fig:phase-fit-2pl} we show respectively the orbital fit and the phase-folded curves for our proposed two-planet solution. We also note that this Neptune-mass planet orbits slightly outside the inner edge of the optimistic circumstellar habitable zone, located at 0.33 au.
            \begin{figure}
                \centering
                \includegraphics[width=\linewidth]{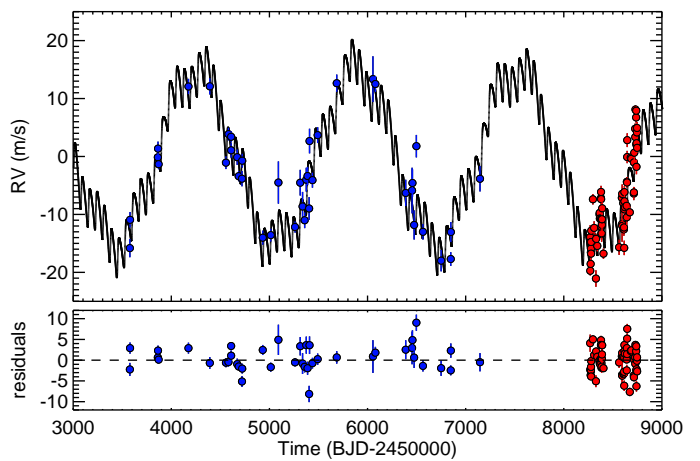}
                \caption{Same as Fig. \ref{fig:orbital-fit-2pl} for the two-planets solution including the year-long circular signal.}
                \label{fig:orbital-fit-3pl}
            \end{figure}
            \begin{figure}
                \centering
                \includegraphics[width=\linewidth]{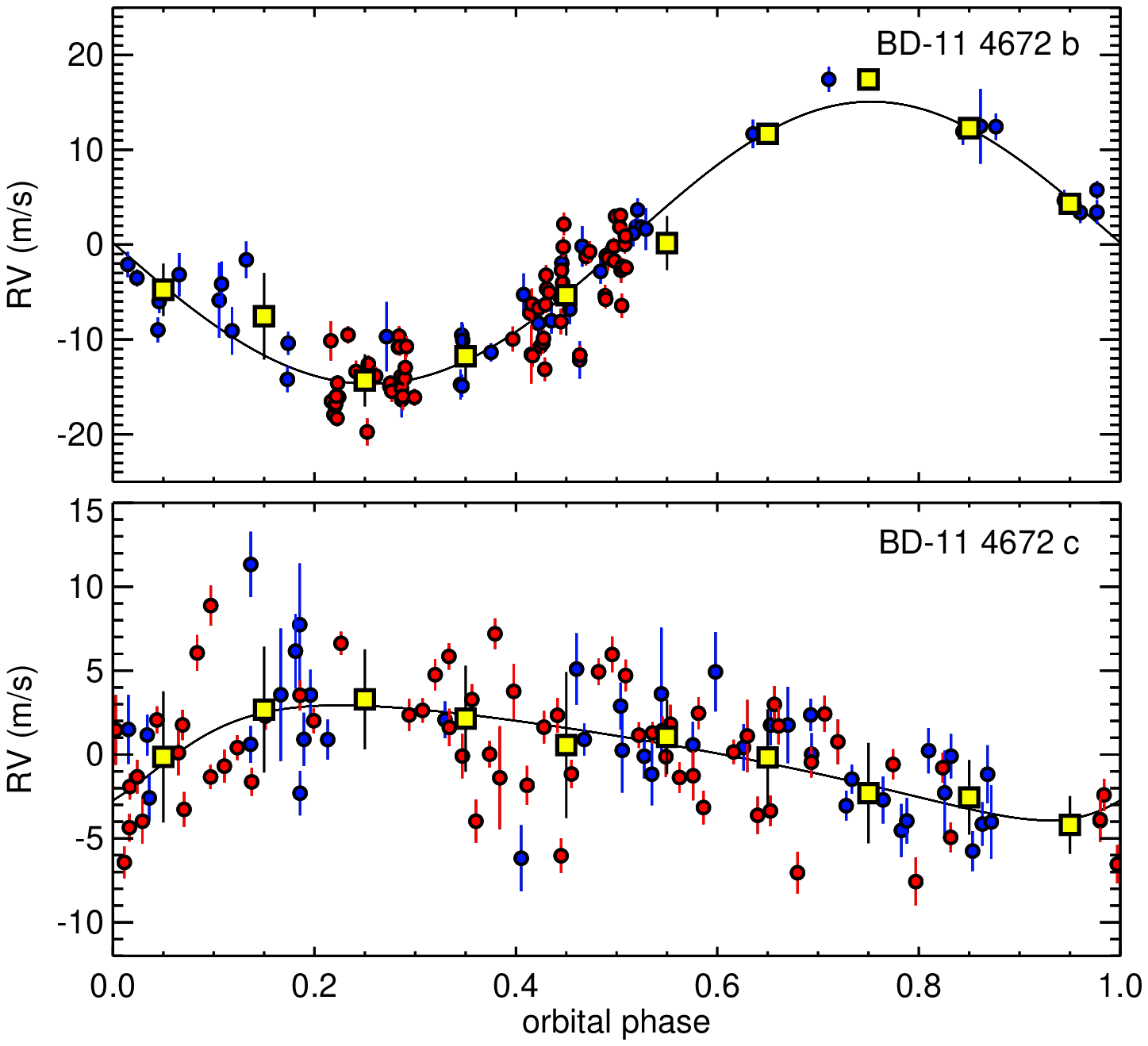}
                \caption{Same as Fig. \ref{fig:phase-fit-2pl} for the two-planets solution including the year-long circular signal}
                \label{fig:phase-fit-3pl}
            \end{figure}
            \begin{table*}
            \caption{Adopted priors and results of the Markov chain Monte Carlo orbital fits. The listed best fit values and uncertainties are the medians and 16$^{\rm th}$ - 84$^{\rm th}$ percentiles of the posterior distributions, respectively.}       \label{table:planets-parameters}
                \centering
                \begin{tabular}{l c c c c}
                \hline\hline
                    Parameter                                       & Priors                         & 1 planet                      & 2 planets                     & 2 planets + year-long signal\\
                                                                    &                                &                               &                               & (adopted solution)\\
                \hline
                    $K_{\rm b}$ ($\rm{m}\,\rm{s}^{-1}$)             & $\mathcal{U}~(0,+\infty)$      & $14.58_{-1.04}^{+1.08}$       & $14.59_{-0.98}^{+1.01}$       & $14.88_{-1.22}^{+1.10}$ \\[3pt]
                    $P_{\rm b}$ (d)                                 & $\mathcal{U}~(1000,2000)$      & $1628\pm12$                   & $1631\pm13$                   & $1634_{-14}^{+13}$ \\[3pt]
                    $\sqrt{e_{\rm b}}\cos{\omega_{\rm b}}$          & $\mathcal{U}~(-1,1)$           & $0.07_{-0.15}^{+0.13}$        & $0.10_{-0.17}^{+0.14}$        & $0.11_{-0.18}^{+0.15}$ \\[3pt]
                    $\sqrt{e_{\rm b}}\sin{\omega_{\rm b}}$          & $\mathcal{U}~(-1,1)$           & $-0.28_{-0.10}^{+0.16}$       & $-0.20_{-0.12}^{+0.17}$       & $-0.04_{-0.14}^{+0.16}$ \\[3pt]
                    $T_{0,b}$ (BJD-2450000)                         & $\mathcal{U}~(6700,8700)$      & $7905.4_{-49.0}^{+42.1}$   & $7908.6_{-52.2}^{+41.9}$   & $7915.3_{-48.3}^{+33.9}$ \\[3pt]
                    $e_{\rm b}$                                     & derived                        & $0.10\pm0.06$                 & $0.08_{-0.05}^{+0.06}$        & $0.05_{-0.03}^{+0.05}$ \\[3pt]
                    $\omega_{\rm b}$ (deg)                          & derived                        & $281.94_{-36.57}^{+28.11}$    & $287.73_{-78.12}^{+35.90}$    & $343.88_{-66.95}^{+92.63}$ \\[3pt]
                    $M_{\rm b}\sin{i}$ ($M_{\rm{Jup}}$)             & derived                        & $0.63\pm0.05$                 & $0.63\pm0.05$                 & $0.65_{-0.06}^{+0.05}$ \\[3pt]
                    $a_{\rm b}$ (au)                                & derived                        & $2.35\pm0.04$                 & $2.36\pm0.04$                 & $2.36\pm0.04$ \\[3pt]
                \hline
                    $K_{\rm c}$ ($\rm{m}\,\rm{s}^{-1}$)             & $\mathcal{U}~(0,+\infty)$      & -                             & $3.48_{-0.67}^{+0.78}$        & $3.42_{-0.64}^{+0.70}$ \\[3pt]
                    $P_{\rm c}$ (d)                                 & $\mathcal{U}~(70,80)$          & -                             & $74.18_{-0.12}^{+0.10}$       & $74.20_{-0.08}^{+0.06}$ \\[3pt]
                    $T_{0,c}$ (BJD-2450000)                         & $\mathcal{U}~(8620,8700)$      & -                             & $8685.4_{-5.9}^{+5.6}$     & $8687.5_{-5.92}^{+5.66}$ \\[3pt]
                    $\sqrt{e_{\rm c}}\cos{\omega_{\rm c}}$          & $\mathcal{U}~(-1,1)$           & -                             & $-0.14_{-0.23}^{+0.28}$       & $-0.24_{-0.20}^{+0.24}$ \\[3pt]
                    $\sqrt{e_{\rm c}}\sin{\omega_{\rm c}}$          & $\mathcal{U}~(-1,1)$           & -                             & $-0.51_{-0.16}^{+0.25}$       & $-0.54_{-0.14}^{+0.21}$ \\[3pt]
                    $e_{\rm c}$                                     & derived                        & -                             & $0.34_{-0.17}^{+0.18}$        & $0.40_{-0.15}^{+0.13}$ \\[3pt]
                    $\omega_{\rm c}$ (deg)                          & derived                        & -                             & $253.68_{-33.98}^{+30.49}$    & $245.92_{-26.58}^{+23.02}$ \\[3pt]
                    $M_{\rm c}\sin{i}$ ($M_{\oplus}$)               & derived                        & -                             & $16.59_{-2.97}^{+3.14}$       & $15.37_{-2.81}^{+2.97}$ \\[3pt]
                    $a_{\rm c}$ (au)                                & derived                        & -                             & $0.30\pm0.01$                 & $0.30\pm0.01$ \\[3pt]
                \hline
                    $K_{\rm yr}$ ($\rm{m}\,\rm{s}^{-1}$)            & $\mathcal{U}~(0,+\infty)$      & -                             & -                             & $2.40_{-0.52}^{+0.54}$ \\[3pt]
                    $P_{\rm yr}$ (d)                                & $\mathcal{U}~(200,400)$        & -                             & -                             & $365.46_{-36.55}^{+5.96}$ \\[3pt]
                    $T_{0,yr}$ (BJD-2450000)                        & $\mathcal{U}~(8300,8700)$      & -                             & -                             & $8480.4_{-20.01}^{+21.81}$ \\[3pt]
                \hline
                    $\gamma_{\rm HARPS}$ ($\rm{m}\,\rm{s}^{-1}$)    & $\mathcal{U}~(-\infty,+\infty)$& $3.83_{-0.64}^{+0.69}$        & $3.96\pm0.62$                 & $3.83_{-0.82}^{+0.87}$ \\[3pt]
                    $\gamma_{\rm HARPS-N}$ ($\rm{m}\,\rm{s}^{-1}$)  & $\mathcal{U}~(-\infty,+\infty)$& $9.78_{-1.23}^{+1.20}$        & $9.59_{-1.27}^{+1.30}$        & $9.11_{-1.33}^{+1.25}$ \\[3pt]
                    Jitter$_{\rm HARPS}$ ($\rm{m}\,\rm{s}^{-1}$)    & $\mathcal{U}~(0,+\infty)$      & $3.21_{-0.47}^{+0.57}$        & $2.58_{-0.48}^{+0.65}$        & $2.42_{-0.47}^{+0.55}$ \\[3pt]
                    Jitter$_{\rm HARPS-N}$ ($\rm{m}\,\rm{s}^{-1}$)  & $\mathcal{U}~(0,+\infty)$      & $4.17_{-0.38}^{+0.44}$        & $3.49_{-0.37}^{+0.40}$        & $2.98_{-0.30}^{+0.34}$ \\[3pt]
                \hline
                    residuals rms ($\rm{m}\,\rm{s}^{-1}$)           &                               & 4.00                          & 3.56                          & 3.01 \\[3pt]
                    maximum $\log{\mathscr{L}}$                     &                               & -199.39                       & -180.18                       & -167.35 \\[3pt]
                    BIC                                             &                               & 440.83                        & 425.78                        & 414.13 \\[3pt]
                \hline
                \end{tabular}
            \end{table*}
            \par As shown in the third panel of Fig. \ref{fig:rv-gls}, the periodogram of the two-planet residuals features a low-significance peak with $\rm{FAP}=5\%$ at 367 d. This one-year residual peak is likely to be a spurious signal induced by the Earth's motion around the Sun, similarly to the annual signal first reported and characterised in \cite{dumusqueetal2015} using HARPS data, in which imperfections in the HARPS CCD can lead to a deformation of specific spectral lines passing by the crossing block stitching of the detector. A similar spurious signal has recently been detected in HARPS-N also, as reported in \cite{benatti2020}. As prescribed in \cite{dumusqueetal2015}, the correction and removal of this low-significance signal is possible either by removing from the correlation mask the spectral lines affected by this effect or by fitting a sinusoidal with a one-year period to the radial velocity timeseries. While the first approach is beyond the scope of this paper and may prove more fruitful for follow-up studies of system BD-11~4672, the inclusion of an additional one-year sine signal to our radial velocity fit may help in provide a more complete and robust characterization of the system.
            \par We find the best fit three-signal model to feature eccentric solutions for planets b and c and a circular solution for the third, yearly signal (see fourth column of Table \ref{table:planets-parameters} and Fig. \ref{fig:orbital-fit-3pl} and \ref{fig:phase-fit-3pl}). In this solution, we find planet b and c to have semiamplitudes $K_{\rm b}=14.88_{-1.22}^{+1.10}$ $\rm{m}\,\rm{s}^{-1}$ and $K_{\rm c}=3.42_{-0.64}^{+0.70}$ $\rm{m}\,\rm{s}^{-1}$, orbital periods $P_{\rm b}=1634_{-14}^{+13}\,\rm{d}$ and $P_{\rm c}=74.20_{-0.08}^{+0.06}\,\rm{d}$, eccentricities $e_{\rm b}=0.05_{-0.03}^{+0.05}$ and $e_{\rm c}=0.40_{-0.15}^{+0.13}$, from which we derive minimum masses $M_{\rm b}\sin{i}=0.65_{-0.06}^{+0.05}\,M_{\rm{Jup}}$, $M_{\rm c}\sin{i}=15.37_{-2.81}^{+2.97}\,M_{\oplus}$ and semimajor axes $a_{\rm b}=2.36\pm0.04\,\rm{au}$, $a_{\rm c}=0.30\pm0.01\,\rm{au}$. The circular Keplerian which we fit the year-long signal with has semiamplitude $K_{\rm yr}=2.40_{-0.52}^{+0.54}$ $\rm{m}\,\rm{s}^{-1}$ and period $P_{\rm c}=365.46_{-36.55}^{+5.96}\,\rm{d}$. The main peak in the three-signal residual periodogram (shown in the fourth panel of Fig. \ref{fig:rv-gls}) has FAP = 11\% and is found at 43 d, a period we identified in Sect. \ref{subsec:activity} as related to stellar rotation or activity, along with its alias peak at 7 d. The low FAP of both peaks and the lack of convergence shown by the tentative four-signals fits we have run suggest that the available radial velocity time series do not currently hint at the presence of additional companions.
            \par All orbital parameters for the two planets in the system are compatible with those returned by the two-Keplerian solution; additionally, with a BIC value of 414.13 and a $\Delta\rm{BIC}=-11.65$ it is clear that the three-signal solution is statistically favoured to the two-Keplerian solution detailed in the previous paragraphs. Finally, we note that the $\Delta\rm{BIC}=-26.40$ between this three-signal solution and the single-planet one represents a much stronger evidence in favour of the presence of planet c that the one provided by the $\Delta\rm{BIC}=-15.05$ between the two-Keplerian and single-planet solutions, and we therefore adopt this orbital solution.

    \section{Dynamics} \label{sec:dynamics}
        As discussed in Sect. \ref{subsec:time series-fit}, we detect the presence of the new Neptune-mass planet BD-11~4672~c, which at semimajor axis $a_{\rm c}=0.30\pm0.01\,\rm{au}$ is located just outside of the inner edge of the optimistic circumstellar habitable zone, which ranges from 0.33 au to 0.81 au.
        \par
\begin{figure}
            \centering
            \includegraphics[width=\linewidth]{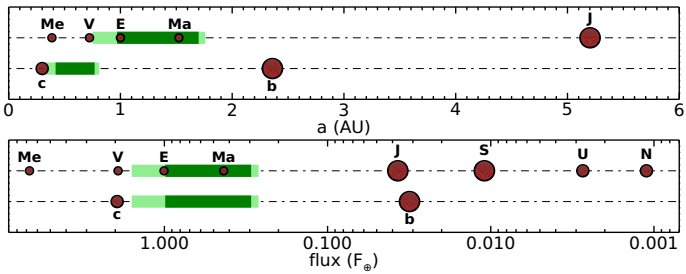}
            \caption{The architecture of planetary system BD-11~4672 compared with that of the Solar System in semi-major axis (top panel) and incident flux scale (bottom panel). The conservative and optimistic habitable zones are shown as dark and light green bands respectively.}
            \label{fig:architecture}
        \end{figure}

 Fig. \ref{fig:architecture} shows the architecture of the BD-11~4672 system as described by our three-signal solution and a comparison with the Solar System. It is interesting to note that planet b receives an incident flux $F_{\rm b}=0.03\,\rm{F}_\oplus$, while planet c has an incident flux of $F_{\rm c}=1.95\,\rm{F}_\oplus$ comparable to that of Venus. We note, however, that the stellar parameters and planetary masses of the BD-11~4672 system are very different from those of the aforementioned Solar System bodies and therefore the analogy between the two systems are only for the sake of comparing the incident fluxes of the planets involved.
        \par It may be interesting to pursue the study of the dynamical evolution and long-term stability of possible lower-mass additional companions in the habitable zone of the system, especially to gauge the dynamical effect of the presence of the newly detected eccentric Neptune-mass planet located just outside the inner edge of the habitable zone. While a full and detailed analysis of dynamical stability of the system lies beyond the scope of this paper, a preliminary and accurate approach to the problem is represented by the angular momentum deficit (AMD) stability criterion presented and detailed in \cite{laskar2000,laskar2017,petit2018}.
        \par With this method, a $n$-body planetary system is considered to be AMD-stable if its angular momentum deficit, defined as the difference between the norm of the angolar momentum of the system and that of a coplanar and circular system having the same semimajor axes $a_{\rm k}$:
        \begin{equation}
            C=\sum_{\rm k=1}^{n} M_{\rm k}\sqrt{G M_\star a_{\rm k}}\left(1-\sqrt{1-e_{\rm k}^2}\cos{i_{\rm k}}\right)
        \end{equation}
        is not sufficient to allow for planetary collisions. In particular, a pair of planets in which the inner one has mass $M$, semimajor axis $a$ and eccentricity $e$ and the outer one has $M^\prime$, $a^\prime$, $e^\prime$ is considered AMD-stable if:
        \begin{equation}
            \beta=\frac{\mathscr{C}}{C_{\rm{cr}}}<1
        \end{equation}
        being $\mathscr{C}$ the relative AMD defined as:
        \begin{equation}
            \mathscr{C}=\gamma\sqrt{\alpha}\left(1-\sqrt{1-e^2}\right)+\left(1-\sqrt{1-e^{\prime 2}}\right)
        \end{equation}
        having denoted $\gamma=M/M^\prime$ and $\alpha=a/a^\prime$, and $C_{\rm{cr}}$ is the critical AMD similarly computed for the pair of critical eccentricities $e_{\rm cr}$, $e_{\rm cr}^\prime$ that satisfy the collision conditions:
        \begin{equation}
            \begin{split}
                & \alpha e_{\rm cr}+ e_{\rm cr}^\prime-1+\alpha=0 \\
                & \alpha e_{\rm cr}+\frac{\gamma e_{\rm cr}}{\sqrt{\alpha\left(1- e_{\rm cr}^2\right)}+\gamma^2 e_{\rm cr}^2}=0
            \end{split}
        \end{equation}
        For multiplanetary systems, the whole system is considered AMD-stable if for every adjacent pair of planets the condition $\beta<1$ is satisfied. While it is important to note that the AMD stability criterion is unable to take into account mean motion resonances and secular chaotic interactions that can further promote or impede the dynamical stability as observed in our own Solar System \citep{laskar2008,laskar2009,batygin2015}, the AMD stability criterion provides a useful test for assessing the long-term stability of planetary systems and has recently been fruitfully used to characterize the dynamical stability of exoplanetary systems and has been often found to be in agreement with Hill stability criterion and dynamical simulations \citep[e.g.][]{zinzi2017,agnew2018,petit2018,stock2020}.
        \par As a first test on the AMD-stability of the BD-11~4672 system, we calculated the $\beta$ of the pair composed by planets b and c as characterized by our best fit orbital solution and assuming coplanarity; in order to also account for the uncertainties on relevant quantities $a_{\rm{b,c}}$, $M_{\rm{b,c}}$ and $e_{\rm{b,c}}$ we performed the AMD-stability test for $10^4$ random values of said quantities uniformily drawn within their best fit errorbars reported in Sect. \ref{subsec:time series-fit} and listed in the fourth column of Table \ref{table:planets-parameters}. We find values of $\beta_{\rm{b,c}}$ ranging from 0.002 to 0.027, proving the long-term stability of our proposed two-planets solution.This is in agreement with a dynamical simulation we have performed with the \texttt{MERCURY} N-body integrator package \citep{chambers1999} using astrocentric input coordinates and the hybrid symplectic/Bulirsch-Stoer integrator with an 1 day initial timestep, also showing that the best fit two-planet configuration is stable up to 10 Myr, as would also be expected by computing the Hill radii of the planets as:
        \begin{equation}
           R_{\rm{H}}=a(1-e)\sqrt[3]{\frac{M_p}{3 M_\star}}
        \end{equation}
        and obtaining $R_{\rm{H,b}}=0.146\,\rm{au}$ and $R_{\rm{H,c}}=0.005\,\rm{au}$, from which it is evident that planets b and c are separated by many mutual Hill radii.
        \par In order to assess the long-term dynamical stability of possible additional planets orbiting within the habitable zone of the system and especially its dependence on the presence of planet c near its inner edge having a significant eccentricity of $e_{\rm c}={0.40_{-0.15}^{+0.13}}$, we inject a single additional planet having dynamical mass $M_{\rm{inj}}$ between 1 and 15 $M_\oplus$, eccentricity $e_{\rm{inj}}$ between 0 and 0.5 and semimajor axis $a_{\rm{inj}}$ between the inner and outer boundary of the optimistic habitable zone (0.33-0.81 au). Assuming coplanarity for all bodies in the system, we calculate the $\beta_{\rm{b,inj}}$ and $\beta_{\rm{c,inj}}$ for each injection, again making $10^4$ random draws of the $a$, $M$ and $e$ of known planets b and c within their best fit uncertainties for each realization of $(M_{\rm{inj}},a_{\rm{inj}})$. Following \cite{laskar2017} the three-planets system is considered AMD-stable if and only if both $\beta_{\rm{b,inj}}$ and $\beta_{\rm{c,inj}}$ are below 1.
        \begin{figure*}
            \centering
            \includegraphics[width=\linewidth]{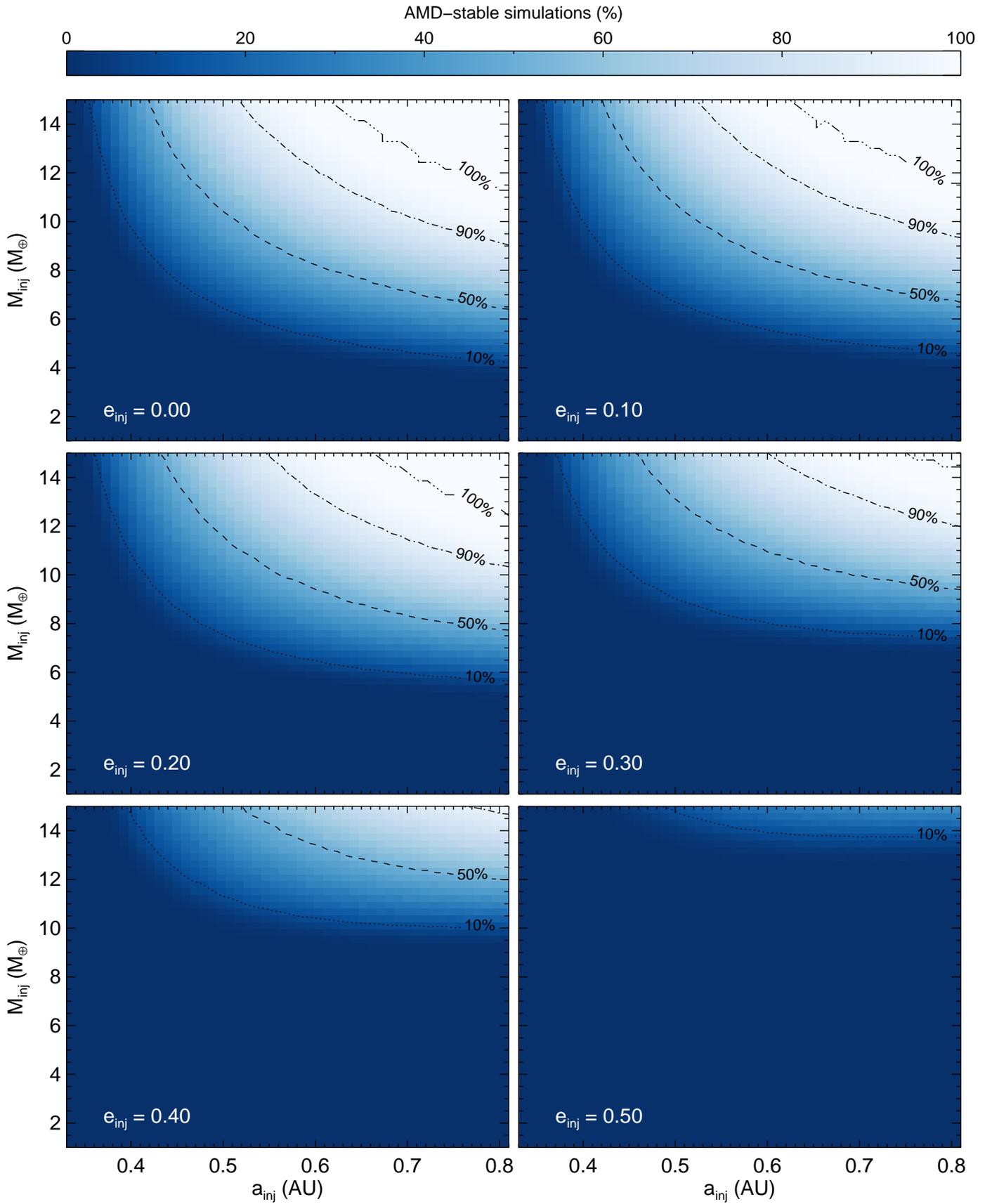}
            \caption{Results of the AMD-stability tests made for the two-planets solution of system BD-11~4672 and injecting a single additional planet in the circumstellar habitable zone, colorcoded according to the fraction of AMD-stable three-planets systems obtained for each injection and $10^4$ random draws of the $a$, $M$ and $e$ of known planets b and c within their best fit uncertainties.}
            \label{fig:amd}
        \end{figure*}
        \par In Fig. \ref{fig:amd} we show the results of the AMD stability tests we performed on a 50x50 $(M_{\rm{inj}},a_{\rm{inj}})$ grid, the color scale representing the fraction of AMD-stable three-planet systems obtained from the $10^4$ random draws of $a_{\rm{b,c}}$, $M_{\rm{b,c}}$ and $e_{\rm{b,c}}$. As a first point, we must note that the more massive and distant planet b has no impact on the AMD-stability of the habitable zone by virtue of its near-zero eccentricity, as every configuration tested returned $\beta_{\rm{b,inj}}<1$; as could be expected from its proximity and eccentricity planet c represents the main influence on the stability of the habitable zone of the system, significantly disrupting the stability of neighbouring planetary bodies.
        \par Indeed, almost every injection of additional planets with $a_{\rm{inj}}<0.4\,\rm{au}$ is found to result in an AMD-unstable system regardless of its $e_{\rm{inj}}$ and $M_{\rm{inj}}$, with less than 30\% of the simulations being AMD-stable only in the $e_{\rm{inj}}<0.3$ and $M_{\rm{inj}}>10\,M_\oplus$ cases; the more eccentric the additional planet is, the less AMD-stable the three-planets system grows, finally resulting in a fully AMD-unstable habitable zone for $e_{\rm{inj}}>0.40$. Ultimately, AMD-stability appears to be possible for more than 90\% of the simulated systems only in the case of Neptune-mass ($M_{\rm{inj}}\geq10$) planets orbiting on relatively low-eccentric ($e_{\rm{inj}}\leq0.30$) orbits in the outer portion ($a_{\rm{inj}}>0.5\,\rm{au}$) of the habitable zone. Barring resonances and secular interaction that can stabilize an AMD-unstable system or viceversa  \citep{zinzi2017,agnew2018,petit2018,stock2020} that are not yet considered by the AMD stability test, we therefore expect yet-undetected temperate Neptune-mass planetary bodies, if any, to be present in the BD-11~4672 system only on low- to moderate-eccentricity orbits on the outer region of the circumstellar habitable zone.
        
    \section{Discussion and conclusions} \label{sec:conclusions}
        In the present work we have reported the results of the high-precision monitoring of the K-dwarf BD-11~4672 conducted with HARPS-N within the GAPS programme from May 2018 to September 2019, performed in order to search additional subgiant planetary companions to the already known giant planet BD-11~4672~b first detected by \cite{moutou2015}.
        \par We provide a new characterization of the physical properties of the host star, finding larger values of mass and radius than those reported in the literature; we also characterize the star as a metal-poor member of the thin-disc stellar population. Additionally, it can be noted that BD-11~4672 appears to be among the most $\alpha$-poor thin-disc stars at similar [Fe/H]. This makes this K-dwarf a possibly interesting counter-example to previous works \citep[e.g][]{haywood2008,haywood2009,kang2011,adibekyan2012a,adibekyan2012c,maldonado2018} that noted a tendency for metal-poor planet-hosting stars to feature an overabundance of $\alpha$-elements possibly promoting the core-accretion formation of gas giant planets by compensating for low [Fe/H]. By studying the stellar activity of BD-11~4672 we find three dominant periodicities in the activity indices Ca~{\sc ii}  H\&K, H$\alpha$, and Na~{\sc i} D$_{\rm 1}$ D$_{\rm 2}$ at $\sim3000$, $\sim200$ and $\sim40$ days. Finally, combining the newly derived projected rotational velocity value of $v\sin{i} = 1 \pm 0.5\,\rm{km}\,\rm{s}^{-1}$, and the empirical relations between stellar rotation and mean activity level, we provide an estimate of rotation period of $P_{\rm{rot}}\sim25$ days. 
        \par The densely sampled time series obtained with HARPS-N allowed us, in conjunction with literature HARPS radial velocity measurements, to provide an updated orbital solution to known planet BD-11~4672~b. Our orbital solution is compatible and better constrained than the one of the discovery paper \citep{moutou2015}, although it is interesting to note that our solution features slightly larger values of minimum mass and semimajor axis for the planet, mainly stemming from the larger stellar mass we obtained from our spectroscopic characterization of the host star. In addition to this, we report the detection of new planet BD-11~4672~c having an orbital period of $74.20_{-0.08}^{+0.06}$ days, eccentricity of $0.40_{-0.15}^{+0.13}$ and radial velocity semiamplitude of $3.42_{-0.64}^{+0.70}$ $\rm{m}\,\rm{s}^{-1}$, for which we derive minimum mass of $15.37_{-2.81}^{+2.97}\,M_\oplus$ and semimajor axis of $0.30\pm0.01\,\rm{au}$. Receiving an incident flux of 1.95 $F_\oplus$, we note that this Neptune-mass planet is found on a slightly more inner location than the optimistic circumstellar habitable zone. We also note that in the historical HARPS time series a low-significance peak is present around 71 d with a high $\rm{FAP}=52\%$ in the single-planet residuals; it is therefore clear that the inclusion of our high-cadence HARPS-N data is instrumental in confirming and identifying this residual peak.
        \par To investigate the effects that Neptune-mass planet c would have on the dynamical stability of additional temperate low-mass planets by virtue of its eccentricity and proximity to the inner edge of the habitable zone of the system, we have conducted AMD stability tests to evaluate the long-term stability of the system with the injection of additional planetary bodies with $M_{\rm{inj}}=1-15\,M_\oplus$ and $e_{\rm{inj}}=0-0.5$ orbiting within the habitable zone. In doing so, we find that planet c has a severely negative effect on the region of parameters space in which the injection of an additional planet keeps the system AMD-stable, only allowing the long-term stability of the more massive, distant and less eccentric injected temperate bodies.
        \begin{figure}
            \centering
            \includegraphics[width=\linewidth]{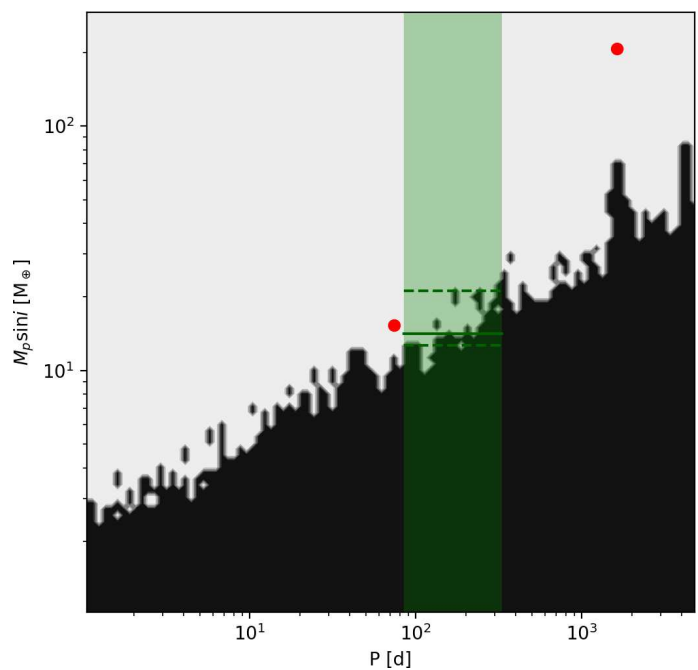}
            \caption{Detection map for the complete RV time series of BD-114672. The white part corresponds to the area in the period - minimum mass space where additional signals could be detected if present in the data, while the black region corresponds to the area where the detection probability is negligible. The red circles mark the position in the parameter space of the two planets in the system. The green areas denote the circumstellar habitable zone, and the horizontal green lines mark the corresponding median detection threshold (solid line), with its 16-84$\%$ uncertainties (dashed lines).}
            \label{fig:detection_lim}
        \end{figure}
        \par We computed the detection threshold for the HARPS-N dataset following the Bayesian approach from \citet{tuomi14}. We applied this technique on the combined RV time series, including in the model the three signals identified in Sect. \ref{subsec:time series-fit}. The 2D-map of the detection function is shown in Fig. \ref{fig:detection_lim}: focusing on the circumstellar habitable zone as studied in our dynamical analysis in Sect. \ref{sec:dynamics}, we obtain a minimum-mass detection threshold of $14.19^{+6.96}_{-1.53}$ M$_\oplus$. Thus it would be possible for a low-eccentricity super-Earth to remain undetected in our data, inside the AMD-stable orbital region shown in Fig. \ref{fig:amd}.
        \par Another source of information potentially useful in characterizing the BD-11~4672 planetary system comes from the use of astrometric measurements, especially from future data releases of the Gaia mission. Indeed, using the relation
            \begin{equation}\label{eq:astrometricsignal}
                \alpha=\frac{M_{\rm p}}{M_\star}\frac{a}{d},
            \end{equation}
        we can provide an estimate of the astrometric signature $\alpha$ produced by each planet in the system and evaluate which astrometric contribution could be in principle detected at high S/N ($\alpha\gtrsim50\,\mu\rm{as}$), using the nominal minimum mass obtained from our best fit solutions to provide a lower estimate of $\alpha$. While the astrometric signature $\alpha_{\rm c}\sim0.8\,\mu\rm{as}$ of Neptune-mass planet c is clearly too low to be detected by Gaia, we find giant planet b to have $\alpha_{\rm{b}}\sim83\,\mu\rm{as}$, a value suggesting the possibility of a future astrometric characterization of the planet's orbit.
            \begin{table}
                \caption{Proper motions for BD-11~4672 and their variations from a linear mean motions as retrieved from the Hipparcos-Gaia Catalog of Accelerations \citep{brandt2018,brandt2019}.}       \label{table:proper-motions}
                \centering
                \begin{tabular}{l c c}
                \hline\hline
                                                                   & Hipparcos          & Gaia DR2 \\
                \hline
                    $\mu_\alpha$ ($\rm{mas}\,\rm{yr}^{-1}$)        & $-286.814\pm1.862$ & $-288.590\pm0.156$ \\[3pt]
                    $\mu_\delta$ ($\rm{mas}\,\rm{yr}^{-1}$)        & $-237.047\pm1.419$ & $-235.560\pm0.145$ \\[3pt]
                    $\Delta\mu_\alpha$ ($\rm{mas}\,\rm{yr}^{-1}$)  & $1.466\pm1.863$    & $-0.309\pm0.169$ \\[3pt]
                    $\Delta\mu_\delta$ ($\rm{mas}\,\rm{yr}^{-1}$)  & $-1.528\pm1.420$   & $-0.042\pm0.153$ \\[3pt]
                \hline
                \end{tabular}
            \end{table}
        \par Additionally, the use of variations in the proper motions measurements of BD-11~4672 between the Hipparcos \citep{hipparcos1997} and Gaia DR2 epochs can provide an useful element in estimating a preliminary constraint on the true mass of planet b. We list in Table \ref{table:proper-motions} the variations $\Delta\mu_{\alpha,\delta}$ from a purely linear stellar mean motion for BD-11~4672 as reported in the Hipparcos-Gaia Catalog of Accelerations \citep[HGCA,][]{brandt2018,brandt2019}, in which the Hipparcos and Gaia catalogues have been cross-calibrated to account for systematics and shift all proper motions in the DR2 reference frame; it is also worth noting that the HGCA uncertainties on Gaia DR2 proper motions are larger than those reported in the DR2 catalogue and listed in Table \ref{table:star-parameters} as a result of the different model of error inflation used in \cite{brandt2018,brandt2019}.
        \par Over the $\sim24\,\rm{yr}$ between the Hipparcos and Gaia measurements BD-11~4672 has deviated very little from a purely linear motion in the plane of the sky, the variation between the two measurement epochs being comparable to zero within $\sim1\sigma$. If we assume that the entirety of this low deviation is due to the orbital motion of the outer giant planet b, we can compute the expected variations in proper motions $\Delta\mu$ that would be caused by different orbital inclinations $i_{\rm{b}}$ and therefore true mass $M_{\rm{b}}$, starting at our best fit nominal value of minimum mass $M_{\rm{b}}\sin{i_{\rm{b}}}=0.65\,M_{\rm{Jup}}$ and finding the critical orbital inclination, and therefore maximum true mass, that would cause a proper motion variation equal to that reported in \cite{brandt2018,brandt2019}. Using the traditional expressions for astrometric motions in the Thiele-Innes element representation \citep{makarov2005,sozzetti2014}, we find that any true mass for planet b above $\sim1.98\,M_{\rm{Jup}}$ would produce astrometric accelerations higher than those reported in \cite{brandt2018,brandt2019}, therefore constraining the true mass of planet b well below the brown dwarf traditional limit of $\sim13\,M_{\rm{Jup}}$. This maximum value for $M_{\rm{b}}$ corresponds to orbital inclination of $i_{\rm{b}}>19\degr$; additionally assuming a coplanar configuration for the BD-11~4672 planetary system, this critical inclination would in turn imply a maximum mass of $M_{\rm{c}}\sim47\,M_\oplus$ for planet c, therefore constraining its true mass to be at most about 2.7 times the mass of Neptune.
        \par The BD-11~4672 system is worthy of additional monitoring and study to better characterize its interesting system architecture and investigate its dynamical origins. With an eccentricity of $0.40_{-0.15}^{+0.13}$, planet BD-11~4672 c is notable for being among the most eccentric Neptune-mass exoplanets found at intermediate orbital period. Indeed, it has been noted from both known transiting and radial velocity detected exoplanetary populations \citep{halbwachs2005,ribas2007,brahm2019,correia2020} that Neptune-mass or -sized exoplanets with $P$ between 50 and 100 d are typically found on lower eccentricities, with notable exceptions such as GJ 96 b \citep{hobson2018} and Kepler-278 c \citep{vaneylen2015} that may hint at some shared eccentricity excitation mechanisms. Additionally, while the different stellar and planetary masses do not allow to properly consider the BD-11~4672 system as a scaled-down Solar System analogue, we note again that planets b and c receive a stellar flux similar to that received by Jupiter and Venus respectively, possibly hinting at some shared dynamical history. It is also worth noting that the BD-11~4672 system architecture, with a long-period giant planet and one inner lower-mass planet, is an example of the planetary architecture some of the authors of this paper searched for around 20 solar-type stars as detailed in \cite{barbato2018}. The BD-11~4672 system joins a small but growing group of Doppler-detected exoplanetary systems around K and M dwarfs showing a clear mass hierarchy, with outer more massive companions and inner lower-mass planets such as GJ 676 A \citep{angladaescude2012b} and GJ 221 \citep{locurto2013}. While such sub-solar dwarfs are noted in the literature to have a lower probability of hosting long-period giant planets than Sun-like stars and are usually host to various super-Earths, the presence of such an architecture around BD-11~4672 could be taken as a positive sign on the variety of planetary systems that formation and migration mechanisms can produce on a large range of stellar types.
    
    \begin{acknowledgements}
        The authors wish to thank the referee, Dr.~T.~Trifonov, for the thorough and useful comments which significantly improved an earlier version of the manuscript.
        The GAPS project acknowledges the support by INAF/Frontiera through the "Progetti Premiali" funding scheme of the Italian Ministry of Education, University, and Research. 
        DB acknowledges financial support from INAF and Agenzia Spaziale Italiana (ASI grant n. 014-025-R.1.2015) for the 2016 PhD fellowship programme of INAF.
        MPi gratefully acknowledges the support from the European Union Seventh Framework Programme (FP7/2007-2013) under Grant Agreement No. 313014 (ETAEARTH).
        We acknowledge financial support from the ASI-INAF agreement n.2018-16-HH.0
        This work has made use of data from the European Space Agency (ESA) mission {\it Gaia} (\url{https://www.cosmos.esa.int/gaia}), processed by the {\it Gaia} Data Processing and Analysis Consortium (DPAC, \url{https://www.cosmos.esa.int/web/gaia/dpac/consortium}). Funding for the DPAC has been provided by national institutions, in particular the institutions participating in the {\it Gaia} Multilateral Agreement.
        DB and MPi also wish to thank A.~Baglio, G.~Storti, G.~Poretti and M.~Massironi for their inspirational work in precision mechanics and advanced technology.
    \end{acknowledgements}
  
  \bibliographystyle{aa}
  \bibliography{ref}
    
    \longtab{
                \begin{longtable}{c c c c c c c c}
                    \caption{\label{table:rv-harps} HARPS measurements for BD-11~4672}\\
                    \hline\hline
                        BJD             & $T_{\rm exp}$         & RV                    & BIS                   & FWHM                  & Ca~{\sc ii}   & H$\alpha$ & Na~{\sc i}\\
                                        & ($\rm{s}$)    & ($\rm{m}\,\rm{s}^{-1}$) & ($\rm{km}\,\rm{s}^{-1}$) & ($\rm{km}\,\rm{s}^{-1}$)  &               &           &           \\
                    \hline
                    \endfirsthead
                    \caption{continued.}\\
                    \hline\hline
                        BJD             & $T_{\rm exp}$         & RV                    & BIS                   & FWHM                  & Ca~{\sc ii}   & H$\alpha$ & Na~{\sc i}\\
                                        & ($\rm{s}$)    & ($\rm{m}\,\rm{s}^{-1}$) & ($\rm{km}\,\rm{s}^{-1}$) & ($\rm{km}\,\rm{s}^{-1}$)  &               &           &           \\
                    \hline
                    \endhead
                    \hline
                    \endfoot
                        $2453577.64$ & $450$ & $-11.98\pm1.58$ & $0.0456$ & $6.1098$ & $0.0362\pm0.0007$ & $0.0488\pm0.0002$ & $0.00380\pm0.00005$ \\ 
                        $2453579.64$ & $450$ & $-7.10\pm1.34$ & $0.0278$ & $6.0924$ & $0.0374\pm0.0006$ & $0.0489\pm0.0002$ & $0.00373\pm0.00004$ \\ 
                        $2453862.82$ & $509$ & $3.69\pm1.38$ & $0.0420$ & $6.1024$ & $0.0351\pm0.0006$ & $0.0485\pm0.0002$ & $0.00381\pm0.00004$ \\ 
                        $2453864.81$ & $450$ & $5.19\pm1.21$ & $0.0320$ & $6.0971$ & $0.0362\pm0.0006$ & $0.0484\pm0.0002$ & $0.00372\pm0.00004$ \\ 
                        $2453870.79$ & $890$ & $2.49\pm0.86$ & $0.0392$ & $6.0923$ & $0.0334\pm0.0004$ & $0.0483\pm0.0001$ & $0.00371\pm0.00003$ \\ 
                        $2454174.90$ & $450$ & $15.91\pm1.32$ & $0.0269$ & $6.0897$ & $0.0350\pm0.0006$ & $0.0482\pm0.0002$ & $0.00364\pm0.00003$ \\ 
                        $2454392.50$ & $450$ & $15.94\pm1.40$ & $0.0339$ & $6.0764$ & $0.0307\pm0.0005$ & $0.0485\pm0.0002$ & $0.00370\pm0.00004$ \\ 
                        $2454556.89$ & $450$ & $2.76\pm1.10$ & $0.0392$ & $6.0695$ & $0.0295\pm0.0004$ & $0.0480\pm0.0001$ & $0.00366\pm0.00003$ \\ 
                        $2454582.83$ & $663$ & $7.69\pm1.09$ & $0.0297$ & $6.0841$ & $0.0318\pm0.0005$ & $0.0481\pm0.0001$ & $0.00366\pm0.00003$ \\ 
                        $2454609.77$ & $541$ & $7.24\pm0.97$ & $0.0303$ & $6.0803$ & $0.0323\pm0.0004$ & $0.0491\pm0.0001$ & $0.00368\pm0.00003$ \\ 
                        $2454609.84$ & $541$ & $4.89\pm1.25$ & $0.0310$ & $6.0852$ & $0.0323\pm0.0005$ & $0.0490\pm0.0001$ & $0.00369\pm0.00003$ \\ 
                        $2454671.71$ & $900$ & $3.71\pm1.33$ & $0.0322$ & $6.0640$ & $0.0289\pm0.0005$ & $0.0484\pm0.0002$ & $0.00362\pm0.00003$ \\ 
                        $2454686.59$ & $676$ & $0.43\pm0.89$ & $0.0317$ & $6.0803$ & $0.0299\pm0.0004$ & $0.0482\pm0.0001$ & $0.00367\pm0.00003$ \\ 
                        $2454720.56$ & $667$ & $0.00\pm1.32$ & $0.0312$ & $6.0705$ & $0.0305\pm0.0005$ & $0.0483\pm0.0002$ & $0.00364\pm0.00003$ \\ 
                        $2454722.60$ & $900$ & $3.10\pm1.19$ & $0.0379$ & $6.0778$ & $0.0291\pm0.0005$ & $0.0483\pm0.0001$ & $0.00370\pm0.00003$ \\ 
                        $2454931.90$ & $450$ & $-10.18\pm1.21$ & $0.0381$ & $6.0770$ & $0.0301\pm0.0005$ & $0.0488\pm0.0001$ & $0.00367\pm0.00003$ \\ 
                        $2455013.71$ & $900$ & $-9.74\pm1.10$ & $0.0407$ & $6.1003$ & $0.0326\pm0.0006$ & $0.0483\pm0.0002$ & $0.00371\pm0.00004$ \\ 
                        $2455091.54$ & $450$ & $-0.67\pm3.64$ & $0.0315$ & $6.1237$ & $0.0418\pm0.0014$ & $0.0489\pm0.0003$ & $0.00364\pm0.00007$ \\ 
                        $2455260.90$ & $900$ & $-8.35\pm0.96$ & $0.0408$ & $6.1089$ & $0.0367\pm0.0005$ & $0.0494\pm0.0001$ & $0.00373\pm0.00003$ \\ 
                        $2455313.83$ & $900$ & $-0.69\pm2.22$ & $0.0378$ & $6.1264$ & $0.0345\pm0.0009$ & $0.0496\pm0.0002$ & $0.00379\pm0.00005$ \\ 
                        $2455337.90$ & $500$ & $-4.80\pm2.51$ & $0.0382$ & $6.1337$ & $0.0324\pm0.0010$ & $0.0491\pm0.0002$ & $0.00371\pm0.00005$ \\ 
                        $2455358.87$ & $900$ & $-7.18\pm1.36$ & $0.0418$ & $6.1345$ & $0.0355\pm0.0007$ & $0.0494\pm0.0002$ & $0.00381\pm0.00004$ \\ 
                        $2455375.72$ & $450$ & $-0.17\pm2.05$ & $0.0273$ & $6.1303$ & $0.0350\pm0.0009$ & $0.0495\pm0.0002$ & $0.00385\pm0.00005$ \\ 
                        $2455388.65$ & $450$ & $0.46\pm1.58$ & $0.0290$ & $6.1349$ & $0.0345\pm0.0007$ & $0.0490\pm0.0002$ & $0.00379\pm0.00004$ \\ 
                        $2455404.66$ & $450$ & $-5.14\pm1.97$ & $0.0310$ & $6.1218$ & $0.0367\pm0.0009$ & $0.0491\pm0.0002$ & $0.00381\pm0.00005$ \\ 
                        $2455408.73$ & $900$ & $6.50\pm2.12$ & $0.0224$ & $6.1333$ & $0.0375\pm0.0009$ & $0.0489\pm0.0002$ & $0.00379\pm0.00005$ \\ 
                        $2455438.63$ & $450$ & $-0.26\pm1.29$ & $0.0360$ & $6.1075$ & $0.0329\pm0.0006$ & $0.0489\pm0.0002$ & $0.00377\pm0.00003$ \\ 
                        $2455491.51$ & $676$ & $7.51\pm1.37$ & $0.0319$ & $6.1142$ & $0.0366\pm0.0007$ & $0.0487\pm0.0002$ & $0.00386\pm0.00004$ \\ 
                        $2455685.94$ & $450$ & $16.47\pm1.50$ & $0.0404$ & $6.1450$ & $0.0377\pm0.0007$ & $0.0491\pm0.0002$ & $0.00385\pm0.00004$ \\ 
                        $2456023.85$ & $900$ & $12.65\pm1.30$ & $0.0332$ & $6.1211$ & $0.0471\pm0.0007$ & $0.0520\pm0.0001$ & $0.00381\pm0.00003$ \\ 
                        $2456023.89$ & $600$ & $11.89\pm1.45$ & $0.0428$ & $6.1238$ & $0.0420\pm0.0008$ & $0.0511\pm0.0002$ & $0.00381\pm0.00004$ \\ 
                        $2456054.76$ & $448$ & $17.18\pm3.95$ & $0.0494$ & $6.1539$ & $0.0383\pm0.0014$ & $0.0497\pm0.0003$ & $0.00381\pm0.00008$ \\ 
                        $2456079.79$ & $600$ & $16.34\pm1.39$ & $0.0380$ & $6.1206$ & $0.0355\pm0.0007$ & $0.0492\pm0.0002$ & $0.00378\pm0.00004$ \\ 
                        $2456388.92$ & $415$ & $-2.46\pm2.28$ & $0.0410$ & $6.1057$ & $0.0370\pm0.0009$ & $0.0486\pm0.0002$ & $0.00369\pm0.00005$ \\ 
                        $2456453.80$ & $800$ & $-2.05\pm3.94$ & $0.0298$ & $6.1220$ & $0.0313\pm0.0014$ & $0.0485\pm0.0003$ & $0.00364\pm0.00008$ \\ 
                        $2456457.79$ & $450$ & $-0.75\pm2.36$ & $0.0409$ & $6.1212$ & $0.0352\pm0.0009$ & $0.0484\pm0.0002$ & $0.00384\pm0.00005$ \\ 
                        $2456474.67$ & $450$ & $-7.99\pm2.51$ & $0.0563$ & $6.1177$ & $0.0321\pm0.0012$ & $0.0483\pm0.0003$ & $0.00378\pm0.00007$ \\ 
                        $2456497.76$ & $900$ & $5.61\pm1.94$ & $0.0283$ & $6.1255$ & $0.0402\pm0.0009$ & $0.0479\pm0.0002$ & $0.00395\pm0.00005$ \\ 
                        $2456564.48$ & $538$ & $-9.17\pm1.35$ & $0.0270$ & $6.0805$ & $0.0280\pm0.0006$ & $0.0484\pm0.0002$ & $0.00365\pm0.00003$ \\ 
                        $2456749.90$ & $500$ & $-14.14\pm1.85$ & $0.0450$ & $6.0983$ & $0.0345\pm0.0009$ & $0.0484\pm0.0002$ & $0.00367\pm0.00005$ \\ 
                        $2456847.73$ & $900$ & $-13.87\pm1.19$ & $0.0298$ & $6.0915$ & $0.0296\pm0.0006$ & $0.0477\pm0.0002$ & $0.00382\pm0.00003$ \\ 
                        $2456848.83$ & $900$ & $-9.21\pm1.73$ & $0.0322$ & $6.1045$ & $0.0309\pm0.0007$ & $0.0480\pm0.0002$ & $0.00381\pm0.00004$ \\ 
                        $2457145.91$ & $600$ & $-0.01\pm2.20$ & $0.0228$ & $6.0728$ & $0.0267\pm0.0010$ & $0.0479\pm0.0003$ & $0.00371\pm0.00006$ \\ 
                \end{longtable}
            }
            
    \longtab{
                \begin{longtable}{c c c c c c c c}
                    \caption{\label{table:rv-harpsn} HARPS-N measurements for BD-11~4672}\\
                    \hline\hline
                        BJD             & $T_{\rm exp}$         & RV                    & BIS                   & FWHM                  & Ca~{\sc ii}   & H$\alpha$ & Na~{\sc i}\\
                                        & ($\rm{s}$)    & ($\rm{m}\,\rm{s}^{-1}$) & ($\rm{km}\,\rm{s}^{-1}$) & ($\rm{km}\,\rm{s}^{-1}$)  &               &           &           \\
                    \hline
                    \endfirsthead
                    \caption{continued.}\\
                    \hline\hline
                        BJD             & $T_{\rm exp}$         & RV                    & BIS                   & FWHM                  & Ca~{\sc ii}   & H$\alpha$ & Na~{\sc i}\\
                                        & ($\rm{s}$)    & ($\rm{m}\,\rm{s}^{-1}$) & ($\rm{km}\,\rm{s}^{-1}$) & ($\rm{km}\,\rm{s}^{-1}$)  &               &           &           \\
                    \hline
                    \endhead
                    \hline
                    \endfoot
                        $2458268.63$ & $1200$ & $-4.81\pm2.08$ & $0.0349$ & $6.1400$ & $0.0314\pm0.0008$ & $0.0485\pm0.0002$ & $0.00397\pm0.00005$ \\ 
                        $2458269.64$ & $1200$ & $-10.61\pm0.83$ & $0.0464$ & $6.1334$ & $0.0324\pm0.0003$ & $0.0487\pm0.0001$ & $0.00384\pm0.00002$ \\ 
                        $2458270.69$ & $1200$ & $-9.78\pm1.66$ & $0.0569$ & $6.1282$ & $0.0286\pm0.0005$ & $0.0491\pm0.0002$ & $0.00428\pm0.00004$ \\ 
                        $2458273.67$ & $1200$ & $-9.43\pm1.03$ & $0.0451$ & $6.1359$ & $0.0324\pm0.0004$ & $0.0486\pm0.0001$ & $0.00372\pm0.00002$ \\ 
                        $2458275.62$ & $900$ & $-7.45\pm0.72$ & $0.0424$ & $6.1409$ & $0.0320\pm0.0003$ & $0.0487\pm0.0001$ & $0.00380\pm0.00002$ \\ 
                        $2458276.64$ & $1200$ & $-6.77\pm0.98$ & $0.0480$ & $6.1413$ & $0.0325\pm0.0004$ & $0.0487\pm0.0001$ & $0.00379\pm0.00003$ \\ 
                        $2458277.59$ & $1200$ & $-5.65\pm0.74$ & $0.0440$ & $6.1462$ & $0.0334\pm0.0003$ & $0.0486\pm0.0001$ & $0.00384\pm0.00002$ \\ 
                        $2458278.65$ & $1200$ & $-7.66\pm0.83$ & $0.0468$ & $6.1462$ & $0.0323\pm0.0003$ & $0.0486\pm0.0001$ & $0.00381\pm0.00002$ \\ 
                        $2458279.66$ & $1200$ & $-3.71\pm0.81$ & $0.0478$ & $6.1477$ & $0.0321\pm0.0003$ & $0.0486\pm0.0001$ & $0.00390\pm0.00002$ \\ 
                        $2458296.57$ & $1200$ & $1.73\pm0.91$ & $0.0530$ & $6.1740$ & $0.0367\pm0.0004$ & $0.0489\pm0.0001$ & $0.00376\pm0.00002$ \\ 
                        $2458309.50$ & $1200$ & $-3.21\pm1.15$ & $0.0498$ & $6.1509$ & $0.0312\pm0.0004$ & $0.0491\pm0.0002$ & $0.00390\pm0.00003$ \\ 
                        $2458327.57$ & $1200$ & $-11.96\pm1.43$ & $0.0443$ & $6.1526$ & $0.0324\pm0.0007$ & $0.0491\pm0.0002$ & $0.00385\pm0.00004$ \\ 
                        $2458329.55$ & $1200$ & $-5.10\pm0.89$ & $0.0461$ & $6.1581$ & $0.0336\pm0.0004$ & $0.0488\pm0.0001$ & $0.00376\pm0.00002$ \\ 
                        $2458341.43$ & $1200$ & $-6.33\pm0.95$ & $0.0465$ & $6.1591$ & $0.0350\pm0.0004$ & $0.0490\pm0.0001$ & $0.00364\pm0.00002$ \\ 
                        $2458361.38$ & $1200$ & $-4.01\pm2.10$ & $0.0543$ & $6.1667$ & $0.0447\pm0.0009$ & $0.0517\pm0.0002$ & $0.00383\pm0.00005$ \\ 
                        $2458364.45$ & $1200$ & $-0.92\pm0.95$ & $0.0507$ & $6.1623$ & $0.0351\pm0.0004$ & $0.0492\pm0.0002$ & $0.00394\pm0.00003$ \\ 
                        $2458365.43$ & $1200$ & $-0.62\pm0.72$ & $0.0506$ & $6.1619$ & $0.0354\pm0.0003$ & $0.0490\pm0.0001$ & $0.00393\pm0.00002$ \\ 
                        $2458367.41$ & $1200$ & $-1.60\pm1.11$ & $0.0420$ & $6.1579$ & $0.0342\pm0.0005$ & $0.0496\pm0.0002$ & $0.00387\pm0.00003$ \\ 
                        $2458378.40$ & $1200$ & $1.94\pm0.78$ & $0.0445$ & $6.1646$ & $0.0363\pm0.0004$ & $0.0490\pm0.0001$ & $0.00381\pm0.00002$ \\ 
                        $2458379.40$ & $1200$ & $2.99\pm1.07$ & $0.0474$ & $6.1706$ & $0.0358\pm0.0004$ & $0.0489\pm0.0001$ & $0.00393\pm0.00003$ \\ 
                        $2458380.39$ & $1200$ & $1.75\pm0.94$ & $0.0482$ & $6.1631$ & $0.0368\pm0.0004$ & $0.0491\pm0.0001$ & $0.00374\pm0.00002$ \\ 
                        $2458381.37$ & $1200$ & $-1.79\pm0.77$ & $0.0428$ & $6.1665$ & $0.0363\pm0.0004$ & $0.0491\pm0.0001$ & $0.00376\pm0.00002$ \\ 
                        $2458382.38$ & $1200$ & $-1.64\pm0.67$ & $0.0478$ & $6.1657$ & $0.0367\pm0.0003$ & $0.0493\pm0.0001$ & $0.00380\pm0.00002$ \\ 
                        $2458383.36$ & $1200$ & $-3.05\pm1.19$ & $0.0480$ & $6.1650$ & $0.0357\pm0.0006$ & $0.0491\pm0.0002$ & $0.00379\pm0.00003$ \\ 
                        $2458384.36$ & $1200$ & $-4.27\pm0.91$ & $0.0485$ & $6.1588$ & $0.0362\pm0.0004$ & $0.0488\pm0.0001$ & $0.00374\pm0.00003$ \\ 
                        $2458385.36$ & $1200$ & $-4.15\pm1.46$ & $0.0480$ & $6.1576$ & $0.0338\pm0.0006$ & $0.0490\pm0.0002$ & $0.00384\pm0.00004$ \\ 
                        $2458388.35$ & $1200$ & $-2.70\pm0.72$ & $0.0434$ & $6.1537$ & $0.0358\pm0.0003$ & $0.0487\pm0.0001$ & $0.00375\pm0.00002$ \\ 
                        $2458389.36$ & $1200$ & $-1.73\pm2.15$ & $0.0532$ & $6.1567$ & $0.0336\pm0.0008$ & $0.0489\pm0.0002$ & $0.00369\pm0.00005$ \\ 
                        $2458391.35$ & $1200$ & $0.17\pm1.08$ & $0.0430$ & $6.1557$ & $0.0364\pm0.0004$ & $0.0490\pm0.0001$ & $0.00373\pm0.00002$ \\ 
                        $2458404.34$ & $1200$ & $-7.66\pm0.89$ & $0.0496$ & $6.1607$ & $0.0345\pm0.0004$ & $0.0493\pm0.0001$ & $0.00386\pm0.00002$ \\ 
                        $2458563.75$ & $1200$ & $-6.56\pm1.31$ & $0.0467$ & $6.1450$ & $0.0345\pm0.0006$ & $0.0491\pm0.0002$ & $0.00378\pm0.00004$ \\ 
                        $2458591.67$ & $1200$ & $1.98\pm0.90$ & $0.0489$ & $6.1710$ & $0.0364\pm0.0005$ & $0.0492\pm0.0001$ & $0.00381\pm0.00003$ \\ 
                        $2458593.72$ & $900$ & $-2.55\pm3.06$ & $0.0491$ & $6.1708$ & $0.0387\pm0.0010$ & $0.0492\pm0.0002$ & $0.00381\pm0.00005$ \\ 
                        $2458594.76$ & $900$ & $2.66\pm1.63$ & $0.0504$ & $6.1744$ & $0.0305\pm0.0004$ & $0.0493\pm0.0002$ & $0.00404\pm0.00003$ \\ 
                        $2458595.74$ & $900$ & $-2.87\pm1.15$ & $0.0521$ & $6.1652$ & $0.0357\pm0.0005$ & $0.0491\pm0.0001$ & $0.00383\pm0.00003$ \\ 
                        $2458605.70$ & $1200$ & $1.06\pm0.83$ & $0.0473$ & $6.1617$ & $0.0368\pm0.0004$ & $0.0491\pm0.0001$ & $0.00384\pm0.00002$ \\ 
                        $2458608.73$ & $1200$ & $-3.30\pm0.99$ & $0.0463$ & $6.1598$ & $0.0350\pm0.0004$ & $0.0488\pm0.0001$ & $0.00391\pm0.00003$ \\ 
                        $2458612.72$ & $1200$ & $-3.47\pm1.11$ & $0.0509$ & $6.1653$ & $0.0365\pm0.0005$ & $0.0489\pm0.0001$ & $0.00390\pm0.00003$ \\ 
                        $2458613.66$ & $1200$ & $-3.14\pm0.91$ & $0.0558$ & $6.1616$ & $0.0378\pm0.0004$ & $0.0492\pm0.0001$ & $0.00370\pm0.00002$ \\ 
                        $2458615.66$ & $900$ & $-6.66\pm1.25$ & $0.0514$ & $6.1590$ & $0.0355\pm0.0005$ & $0.0488\pm0.0001$ & $0.00382\pm0.00003$ \\ 
                        $2458616.68$ & $900$ & $0.00\pm0.90$ & $0.0498$ & $6.1599$ & $0.0368\pm0.0004$ & $0.0491\pm0.0001$ & $0.00374\pm0.00002$ \\ 
                        $2458617.66$ & $900$ & $2.96\pm1.07$ & $0.0447$ & $6.1570$ & $0.0361\pm0.0005$ & $0.0486\pm0.0001$ & $0.00377\pm0.00003$ \\ 
                        $2458618.63$ & $900$ & $1.37\pm1.33$ & $0.0468$ & $6.1538$ & $0.0351\pm0.0006$ & $0.0489\pm0.0002$ & $0.00384\pm0.00003$ \\ 
                        $2458622.70$ & $1200$ & $0.36\pm0.90$ & $0.0408$ & $6.1731$ & $0.0365\pm0.0004$ & $0.0488\pm0.0001$ & $0.00382\pm0.00002$ \\ 
                        $2458640.68$ & $1200$ & $0.60\pm0.77$ & $0.0473$ & $6.1608$ & $0.0372\pm0.0004$ & $0.0487\pm0.0001$ & $0.00372\pm0.00002$ \\ 
                        $2458641.60$ & $1200$ & $-1.39\pm1.33$ & $0.0471$ & $6.1577$ & $0.0377\pm0.0006$ & $0.0485\pm0.0001$ & $0.00373\pm0.00003$ \\ 
                        $2458642.67$ & $1200$ & $4.75\pm0.80$ & $0.0470$ & $6.1644$ & $0.0370\pm0.0004$ & $0.0485\pm0.0001$ & $0.00383\pm0.00002$ \\ 
                        $2458644.56$ & $1200$ & $4.63\pm0.90$ & $0.0462$ & $6.1681$ & $0.0356\pm0.0005$ & $0.0485\pm0.0001$ & $0.00381\pm0.00003$ \\ 
                        $2458645.64$ & $1200$ & $9.02\pm1.06$ & $0.0458$ & $6.1633$ & $0.0376\pm0.0005$ & $0.0486\pm0.0001$ & $0.00379\pm0.00003$ \\ 
                        $2458646.64$ & $1200$ & $11.93\pm1.19$ & $0.0480$ & $6.1773$ & $0.0379\pm0.0005$ & $0.0487\pm0.0001$ & $0.00379\pm0.00003$ \\ 
                        $2458672.44$ & $1200$ & $-0.53\pm1.02$ & $0.0512$ & $6.1471$ & $0.0331\pm0.0005$ & $0.0486\pm0.0002$ & $0.00387\pm0.00003$ \\ 
                        $2458682.57$ & $1200$ & $8.93\pm0.95$ & $0.0457$ & $6.1585$ & $0.0373\pm0.0004$ & $0.0486\pm0.0001$ & $0.00384\pm0.00002$ \\ 
                        $2458688.46$ & $1200$ & $8.73\pm1.10$ & $0.0515$ & $6.1752$ & $0.0384\pm0.0004$ & $0.0485\pm0.0001$ & $0.00392\pm0.00003$ \\ 
                        $2458713.42$ & $1200$ & $2.71\pm1.14$ & $0.0442$ & $6.1488$ & $0.0357\pm0.0006$ & $0.0492\pm0.0001$ & $0.00376\pm0.00003$ \\ 
                        $2458714.47$ & $1200$ & $2.91\pm0.94$ & $0.0441$ & $6.1473$ & $0.0341\pm0.0004$ & $0.0491\pm0.0001$ & $0.00382\pm0.00003$ \\ 
                        $2458715.43$ & $1200$ & $8.09\pm1.00$ & $0.0441$ & $6.1489$ & $0.0342\pm0.0004$ & $0.0487\pm0.0001$ & $0.00375\pm0.00003$ \\ 
                        $2458718.46$ & $1200$ & $9.77\pm1.32$ & $0.0407$ & $6.1623$ & $0.0362\pm0.0006$ & $0.0492\pm0.0001$ & $0.00374\pm0.00003$ \\ 
                        $2458727.42$ & $1200$ & $13.90\pm0.90$ & $0.0492$ & $6.1843$ & $0.0400\pm0.0004$ & $0.0499\pm0.0001$ & $0.00400\pm0.00003$ \\ 
                        $2458728.43$ & $1200$ & $12.46\pm0.73$ & $0.0570$ & $6.1869$ & $0.0386\pm0.0003$ & $0.0493\pm0.0001$ & $0.00393\pm0.00002$ \\ 
                        $2458730.43$ & $1200$ & $17.22\pm0.69$ & $0.0511$ & $6.1883$ & $0.0377\pm0.0004$ & $0.0493\pm0.0001$ & $0.00401\pm0.00002$ \\ 
                        $2458737.37$ & $1200$ & $15.85\pm0.93$ & $0.0458$ & $6.1993$ & $0.0402\pm0.0004$ & $0.0499\pm0.0001$ & $0.00382\pm0.00002$ \\ 
                        $2458738.38$ & $1200$ & $17.02\pm0.78$ & $0.0519$ & $6.1928$ & $0.0391\pm0.0003$ & $0.0496\pm0.0001$ & $0.00392\pm0.00002$ \\ 
                        $2458739.37$ & $1200$ & $11.15\pm1.33$ & $0.0427$ & $6.1979$ & $0.0389\pm0.0006$ & $0.0494\pm0.0002$ & $0.00378\pm0.00003$ \\ 
                        $2458740.37$ & $1200$ & $7.33\pm1.28$ & $0.0548$ & $6.1962$ & $0.0390\pm0.0006$ & $0.0494\pm0.0002$ & $0.00379\pm0.00004$ \\ 
                        $2458741.38$ & $1200$ & $11.39\pm0.79$ & $0.0552$ & $6.1899$ & $0.0395\pm0.0004$ & $0.0497\pm0.0001$ & $0.00396\pm0.00002$ \\ 
                        $2458745.38$ & $1071$ & $13.27\pm0.96$ & $0.0505$ & $6.1780$ & $0.0385\pm0.0005$ & $0.0489\pm0.0001$ & $0.00371\pm0.00002$ \\ 
                        $2458746.38$ & $1200$ & $14.04\pm1.04$ & $0.0488$ & $6.1650$ & $0.0382\pm0.0005$ & $0.0491\pm0.0001$ & $0.00385\pm0.00003$ \\ 
                        $2458747.41$ & $1101$ & $10.60\pm0.84$ & $0.0519$ & $6.1652$ & $0.0375\pm0.0004$ & $0.0489\pm0.0001$ & $0.00380\pm0.00002$ \\ 
                \end{longtable}
            }

\end{document}